\documentclass[10pt,journal,compsoc]{IEEEtran}

\usepackage{amssymb}
\usepackage{graphicx}
\usepackage{subfigure}
\usepackage{epstopdf}
\usepackage{gensymb}
\usepackage{color}
\usepackage{amsmath}
\usepackage{bm}
\usepackage{etoolbox}
\usepackage{cite}
\usepackage{array}
\usepackage{booktabs}
\usepackage{multirow}
\usepackage{comment}

\usepackage{threeparttable} 
\usepackage{mathtools}
\usepackage{caption}
\usepackage{pifont}
\definecolor{garrisonpink1}{rgb}{0.858, 0.188, 0.478}

\definecolor{PangLHpink2}{rgb}{0, 0, 1}

\definecolor{yjr}{rgb}{0.975, 0.375, 0}

\definecolor{R}{RGB}{0,150,0}
\definecolor{A}{RGB}{20,20,20}
\definecolor{T}{RGB}{0,0,150}

\usepackage{minibox}
\newcounter{observcntr}
\newcommand*{\observ}[1]{%
    \stepcounter{observcntr}%
    \begin{center}
    \vspace{-4pt}
    \minibox[frame, rule=1pt,pad=3pt]{
        \begin{minipage}[t]{0.95\columnwidth}
        \textbf{Observation~\arabic{observcntr}:} \textit{#1}.
        \end{minipage}
    }
    \vspace{-4pt}
    \end{center}
}

\IEEEoverridecommandlockouts

\begin{document}

\title{Systematically Evaluation of Challenge Obfuscated APUFs}

\author{
Yansong Gao, Jianrong Yao, Lihui Pang, Zhi Zhang, Anmin Fu, Naixue Xiong, Hyoungshick Kim


\IEEEcompsocitemizethanks{\IEEEcompsocthanksitem Y. Gao, J. Yao and A. Fu are with School of Computer Science and Engineering, Nanjing University of Science and Technology, China. e-mail: \{yansong.gao,120106222744,fuam\}@njust.edu.cn}

\IEEEcompsocitemizethanks{\IEEEcompsocthanksitem L. Pang is with School of Electrical Engineering, University of South China, China, and Department of Software, Sungkyunkwan University, South Korea. e-mail: sunshine.plh@hotmail.com}

\IEEEcompsocitemizethanks{\IEEEcompsocthanksitem Z. Zhang is with Data61, CSIRO, Sydney, Australia. e-mail: zhi.zhang@data61.csiro.au}

\IEEEcompsocitemizethanks{\IEEEcompsocthanksitem N. Xue is with National Engineering Research Center for E-Learning, Central China Normal University, Wuhan 430079, China. e-mail: nicholas.xiong@ccnu.edu.cn}

\IEEEcompsocitemizethanks{\IEEEcompsocthanksitem H. Kim is with Department of Software, Sungkyunkwan University, South Korea. e-mail: hyoung@skku.edu}


}

\IEEEpeerreviewmaketitle

\IEEEtitleabstractindextext{		
\begin{abstract}
As a well-known physical unclonable function that can provide huge number of challenge response pairs (CRP) with a compact design and fully compatibility with current electronic fabrication process, the arbiter PUF (APUF) has attracted great attention. To improve its resilience against modeling attacks, many APUF variants have been proposed so far. Though the modeling resilience of response obfuscated APUF variants such as XOR-APUF and lightweight secure APUF (LSPUF) have been well studied, the challenge obfuscated APUFs (CO-APUFs) such as feed-forward APUF (FF-APUF), and XOR-FF-APUF are less elucidated, especially, with the deep learning (DL) methods. This work systematically evaluates five CO-APUFs including three influential designs of FF-APUF, XOR-FF-APUF, iPUF, one very recently design (dubbed as $Mn_{S_{1},S_{2},S_{3}}$-APUF) and our newly optimized design (dubbed as OAX-FF-APUF), in terms of their \textit{reliability, uniformity (related to uniqueness), and modeling resilience}. Three DL techniques of GRU, TCN and MLP are employed to examine these CO-APUFs' modeling resilience---the first two are \textit{newly} explored. With computation resource of a \textit{common personal computer}, we show that all five CO-APUFs with \textit{relatively large scale} can be successfully modeled---attacking accuracy higher or close to its reliability. The hyper-parameter tuning of DL technique is crucial for implementing efficient attacks. Increasing the scale of the CO-APUF is validated to be able to improve the resilience but should be done with minimizing the reliability degradation.  As the powerful capability of DL technique affirmed by us, we recommend the DL, specifically the MLP technique always demonstrating best efficacy, to be always considered for examining the modeling resilience when newly composited APUFs are devised or to a large extent, other strong PUFs are constructed.  
\end{abstract}

\begin{IEEEkeywords}
Challenge obfuscated APUF, FF-APUF, Modeling attack, Deep learning.
\end{IEEEkeywords}
}

\maketitle

\section{Introduction}
The physical unclonable functions (PUF) is alike 'hardware fingerprint' per hardware instance~\cite{gao2020physical}. It exploits uncontrollable fabrication variations induced randomness to extract unique fingerprints. Therefore, no two identical PUF instances can be forged even with same design and same fabrication processes. Given an input (challenge) fed into the PUF, an instance dependent output (response) is produced. According to the number of challenge response pairs (CRPs) yielded, the PUF can be generally categorized into weak PUF and strong PUF~\cite{herder2014physical}. The weak PUF is mainly used for cryptographic key provision, where the volatile key is only derived on demand and then erased after usage, eradicating the secure non-volatile key storage that usually requires additional fabrication steps~\cite{gao2018lightweight}.  Representative weak PUFs are memory-based PUFs~\cite{gao2021noisfre}, especially SRAM PUFs that are intrinsic PUFs for most electronic commodities as the SRAM memories are pervasively used~\cite{holcomb2008power,gao2019building}. The secure and lightweight key provision based on memory PUF is very attracting for low-cost Internet of Things~\cite{gao2018lightweight,gao2021noisfre}. Compared to the weak PUF, the strong PUF has a wider range of applications beyond cryptographic key provision, ranging from lightweight identification, lightweight authentication to advanced cryptographic protocols such as oblivious transfer, bit commitment multi-party computation, and virtual proof of reality~\cite{brzuska2011physically,ruhrmair2012practical,ruhrmair2013practical,damgaard2013unconditionally,ruhrmair2013pufs,ruhrmair2015virtual}. In the IoT era, the most attractive application of the strong PUF is lightweight authentication, especially when the silicon strong PUF is realized.

The most studied silicon strong PUF candidate is the arbiter APUF (APUF), a representative time delay based PUF design, which yields exponential number of CPRs with compact and easy-to-fabrication advantages~\cite{gassend2002silicon}. However, it has been well-known that the basic APUF is susceptible to modeling attacks, where the built (software/mathematical) model can accurately predict the response given an unseen challenge~\cite{lim2005extracting}. The model is normally learned through machine learning techniques with a set of known CRPs as training data. To combat the modeling attacks, various APUF variants have been proposed. APUF variants adding non-linearity into the PUF design to improve their modeling resilience. In this context, the most used means is obfuscating the APUF responses, especially through the XOR, such as XOR-APUF and lightweight secure APUF (LSPUF)~\cite{majzoobi2008lightweight}. MUXPUF~\cite{sahoo2017multiplexer} also obfuscates the APUF response via multiplexing. Modeling resilience of these APUF variants have been extensively studied, which still confronts hardness to resist evolving modeling attacks such as customized logistic regression~\cite{ruhrmair2010modeling,ruhrmair2013puf}, CMA-ES~\cite{becker2015gap}, reliability based modeling attack~\cite{becker2015pitfalls}, and recent deep learning techniques~\cite{aseeri2018machine,santikellur2019deep}. 

As a distinct means of adding non-linearity, obfuscating the APUF challenge has demonstrated promising modeling resilience~\cite{avvaru2020homogeneous}. The most well-known design is feed-forward APUF~\cite{lim2005extracting,ruhrmair2013puf}, FF-APUF, which utilizes an intermediate arbiter insert within the APUF itself to generate a control bit serving as an obfuscated challenge bit (detailed in Section~\ref{sec:preliminary}). The FF-APUF responses can also be XORed to form XOR-FF-APUFs. Recently, there are designs leveraging standalone auxiliary APUF's response(s) to act as obfuscated challenge bit(s), which include the iPUF~\cite{nguyen2019interpose} and $Mn_{S_{1},S_{2},S_{3}}$-APUF~\cite{ebrahimabadi2021novel} (both designs are detailed in Section~\ref{sec:preliminary}). These challenge obfuscated APUFs (termed as CO-APUFs) exhibit greatly enhanced modeling resilience to conventional modeling attacks such as logistic regression (LR), and CMA-ES. However, their modeling resilience against recent DL attacks are less elucidated and understood (see related work in Section~\ref{sec:related}).  

This work aims to systematically evaluate the modeling resilience of five CO-APUFs with a number of popular DL techniques. In addition, we systematically evaluate the uniqueness and reliability performance of each CO-APUFs. The evaluations are performed with unified experimental settings to provide fair comparative analysis. Our main contributions and found results are summarized as below.
\begin{itemize}
    \item We systematically evaluate the performance of up to five CO-APUFs. For each CO-APUF, its reliability, uniformity and modeling resilience are extensively evaluated with a unified experimental setting.
    \item We, for the first time, show the practicality of attacking FF-APUF with recurrent neural networks. Among three specific employed DL techniques including GRU, TCN and MLP, the MLP exhibits the best attacking performance for all five CO-APUFs.
    \item We debunk the security claim of a recent proposed $Mn_{S_{1},S_{2},S_{3}}$-APUF. At the same time, \textit{our configured MLP} can successfully break FF-APUF, XOR-FF-APUF, and iPUF to a larger scale that have not tested by previous studies with less computation resource.
    \item We propose a new optimized design OAX-FF-APUF, which demonstrates improved reliability while retaining its modeling resilience being comparable to the XOR-FF-APUF given same number of underlying FF-APUFs used. So that the OAX-FF-APUF can be an alternative to XOR-FF-APUF when flexibly increasing the FF-APUFs scale to be resilient to modeling attacks.
\end{itemize}

The rest of the paper is organized as follows. Section~\ref{sec:related} presents related work. The focused five CO-APUFs are described in Section~\ref{sec:preliminary}. Three DL techniques and how to mount them on attacking CO-APUFs are detailed in Section~\ref{sec:MLM}. Reliability and uniformity performance per CO-APUF are evaluated in Section~\ref{sec:Exp}, followed by modeling resilience evaluations in Section~\ref{sec:Modeling Resilience}. Further discussions including future work are presented in Section~\ref{sec:discussion}. This work is concluded in Section~\ref{sec:conclusion}.
\section{Related Work}\label{sec:related}

\subsection{Conventional Machine Learning Attack}
With conventional machine learning techniques such as SVM, LR, CMA-ES, there are two influential attacking strategies to evaluate the modeling resilience of silicon strong PUFs, mostly evaluated on APUF variants, especially XOR-APUFs. These two are from R{\"u}hrmair~\cite{ruhrmair2010modeling,ruhrmair2013puf} and Becker~\cite{becker2015gap,becker2015pitfalls}, respectively. One major improvement of Becker's attack over R{\"u}hrmair's is to exploit certain easy-to-obtain side-channel information, e.g., response reliability information~\cite{becker2015gap} or Hamming weight~\cite{becker2015pitfalls}, which significantly reduces the required number of CRPs to gain an accurate trained model and which further removes the requirement for a direct challenge and response relationship. In addition, building upon the reliability based modeling attack, Becker takes advantage of a divide-and-conquer strategy to attack XOR-APUFs~\cite{becker2015pitfalls}. Each underlying APUF is broken one-by-one, reducing the attacking complexity to be linear as a function of $z$ when attacking $z$-XOR-APUFs.

\subsection{Deep Learning Attack} Recently, the DL has been recently explored to attack strong PUF, in particular, APUF variants. In this category, the most powerful technique is MLP. In 2018, Aseeri \textit{et al.}~\cite{aseeri2018machine} were the first to attack $z$-XOR-APUFs of large-scale (i.e. 64-bit $8$-XOR-APUF and 128-bit $7$-XOR-APUF) using neural networks. The MLP here has 3 hidden layers with $2^{z}$ neurons and used \textsf{relu} as the activation function of the hidden layer. Mursi\textit{ et al.}~\cite{mursi2020fast} very recently exploited a three-hidden layer MLP structure for successfully attacking $z$-XOR-APUFs. Nils Wisiol\textit{ et al.}~\cite{wisiol2021neural} successfully broke 64-bit 11-XOR-APUF with 325 million CRPs when reproducing the MLP. Here, the MLP has 3 hidden layers and the number of neurons in the first and third hidden layers is $2^{z-1}$, while the second hidden layer has $2^{z}$ neurons. The activation function of the hidden layer is \textsf{tanh}. All above focuses are XOR-APUFs, which are distinct from our focused CO-APUFs. 

In 2017, Alkatheiri~\textit{et al.}~\cite{alkatheiri2017towards} used a 3-layer (in particular, only one hidden layer) MLP to attack FF-APUFs---but specific configurations of feed-forward loops are missed in~\cite{alkatheiri2017towards}.  Santikellur \textit{et al.}~\cite{santikellur2019deep} evaluated XOR-APUF and its variants, as well as other response obfuscated MUXPUFs~\cite{sahoo2017multiplexer}, and LSPUFs~\cite{majzoobi2008lightweight}---they have also evaluated one of CO-APUFs that is iPUF, which demonstrated the powerful of MLP on modeling them. As for the iPUF, the largest scale is ($4,4$)-iPUF. We have now demonstrated that ($5,5$)-iPUF and ($1,7$)-iPUF are breakable using personal computing resources. Specifically, in~\cite{santikellur2019deep}, the MLP used to attack 64-bit $z$-XOR-APUF has 2 hidden layers when $z\leq 5$, 5 hidden layers when $z=6$. The MLP used to attack 64-bit LSPUF has 4 or 5 hidden layers. When attacking MUXPUF and its variants, the number of hidden layers is 2 or 3 or 4. The MLP with 3 hidden layers was used to attack iPUF in~\cite{santikellur2019deep}. It should be noted that the activation function of the hidden layer is \textsf{relu} in~\cite{santikellur2019deep}---\cite{wisiol2021neural} also uses \textsf{relu}. In our focused CO-APUFs evaluations, we use \textsf{tanh} as it allows usage of negative value, benefiting network training~\cite{wisiol2021neural}. Recently, Avvaru \textit{et al.}~\cite{avvaru2020homogeneous} used MLP to evaluate the XOR-FF-APUF, however, the FF-APUF stage is only 32, which appears to be a uncommon used smaller stage number---normally 64-stage is used. 

Nils\textit{ et al.}~\cite{wisiol2021neural} used enhanced logistic regression (LR) as a baseline and compared the results of two MLP attacks when mounting on XOR-APUFs. It is shown that MLP structure proposed by Mursi\textit{et al.}~\cite{mursi2020fast} performs better than that proposed by Aseeri \textit{et al.}~\cite{aseeri2018machine}. Compared with the enhanced LR---in particular, the activation function uses \textsf{tanh}, the MLP structure proposed by Mursi\textit{et al.}~\cite{mursi2020fast} has lower data complexity (lower number of CRPs for training) when attacking large XOR-APUFs. However, when attacking small XOR-APUFs, the improved logistic regression has more advantages in data complexity~\cite{wisiol2021neural}. 

In summary, the DL techniques, in particular, powerful MLP, have been recently explored to examine the modeling resilience of APUF variants, but mainly on the response obfuscated APUFs such as the XOR-APUF. In contrast, we focus on a range of CO-APUFs, which was believed to have improved modeling resilience compared to XOR-APUFs. In addition, we have considered two new DL techniques including GRU and TCN.

\section{Obfuscated Challenge APUFs}\label{sec:preliminary}
We introduce five CO-APUFs studied in this work. Note that the OAX-FF-APUF is newly proposed by us.
\subsection{Arbiter-PUF}
\begin{figure}[h]
	\centering
	\includegraphics[trim=0 0 0 0,clip,width=0.50\textwidth]{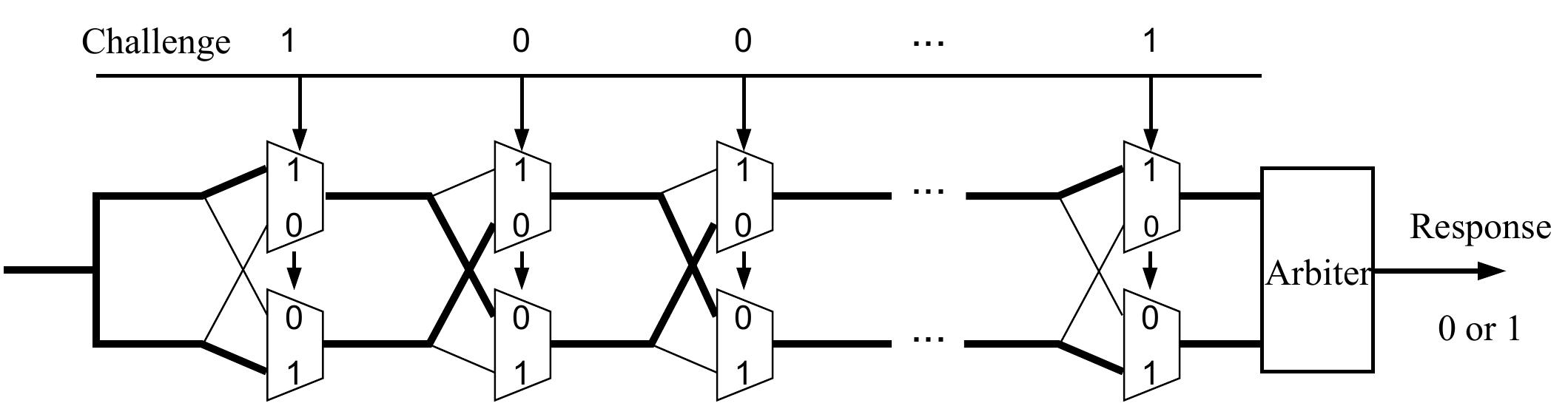}
	\caption{Overview of APUF}
	\label{fig:Arbiter PUF}
\end{figure}

Arbiter-PUF (APUF) is a typical representative of time delay based strong PUF candidate. Fig.~\ref{fig:Arbiter PUF} shows the structure of an APUF. The APUF consists of two parallel signals, which races against each other within $n$ electronic components---2-to-1 multiplexers. Finally, the arbiter determines which signal arrives first and outputs the final response `0' or `1'. 

According to~\cite{lim2005extracting}, APUF can be expressed by a linear additive delay model:
 \begin{equation}
     \Phi[n]=1,\Phi[i]=\prod_{j=i}^{n-1}{(1-2c[j])},i=0,...,n-1, \label{con:feature vector}
  \end{equation}

\begin{equation}
      \Delta={\vec{w}^{T}\vec{\Phi}}, \label{con:delta} 
\end{equation}
where $\vec{w}$ is the weight vector that models the time delay segments in the APUF, $\vec{\Phi}$ is the parity (or generally feature) vector that can be understood as a transformation of the challenge. The dimension of both $\vec{w}$ and $\vec{\Phi}$ is $n+1$ given an $n$-stage APUF.

When given a challenge \textbf{c}, the response $r$ is determined by Eq.~\ref{con:delta}:

\begin{equation}
    r=\begin{cases}
     1, \text{ if } \Delta<0 \\ 
     0, \text{otherwise}.
 \end{cases}
 \label{con:responses}
 \end{equation}
 
\subsection{Feed Forward Arbiter-PUF}
\begin{figure}[h]
	\centering
	\includegraphics[trim=0 0 0 0,clip,width=0.50\textwidth]{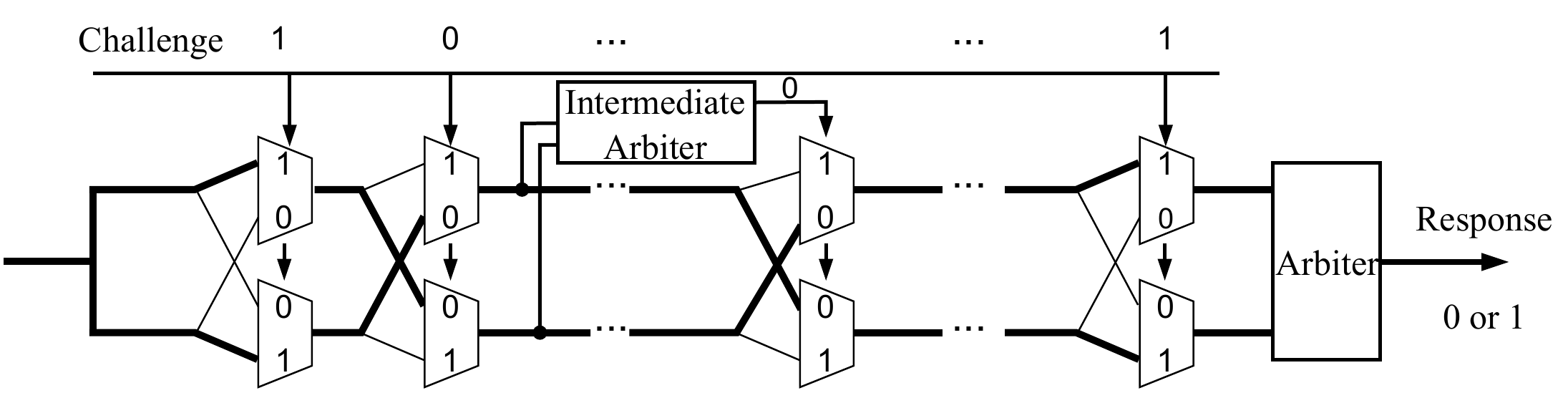}
	\caption{A FF-APUF with one loop.}
	\label{fig:ff-puf with one loop}
\end{figure}

The Feed Forward Arbiter-PUF (FF-APUF)~\cite{lim2005extracting} adds one or more intermediate arbiters within a basic APUF, and the output response of the intermediate arbiter replaces one or multiple bits of the challenge. This is a typical design of obfuscating the APUF challenge bit(s). The structure of a FF-APUF with one loop is depicted in Fig.~\ref{fig:ff-puf with one loop}.

Alkatheiri \textit{et al.}~\cite{alkatheiri2017towards} proposed an improved mathematical model of $n$-stage FF-APUF with one loop. This model can be extended to suit multiple loops. If the FF-APUF has $k$ loops, the math model has $2^{k}$ possibilities. 

As for a single loop to simplify description, if the loop starts at stage $i_{1}$ and ends at stage $i_{2}$, then the response of the FF-APUF can be
modeled by
\begin{equation}
\begin{split}
     &r=sgn(v(n)+\sum_{i\neq i_{2}}\Phi(i)w(i)+w(i_{2})) ~or\\ &r=sgn(v(n)+\sum_{i\neq i_{2}}\Phi(i)w(i)-w(i_{2})),
\end{split}
 \label{con:FF-PUF responses}
 \end{equation}
where $sgn(\cdot)$ is the sign function, $v(n)$ and $w(i)$ are parameters quantifying the difference between signal delays of two paths at the $i_{\rm th}$ stage. Eq.~\ref{con:FF-PUF responses} means that only one of the two equations correctly describes the relationship between the challenge and the response. When the model is extended to FF-APUF with $k$ loops, the model will have $2^{k}$ possibilities but only one of them will have the correct relationship between the response and the challenge bits~\cite{alkatheiri2017towards}.

\begin{figure}[h]
	\centering
	\includegraphics[trim=0 0 0 0,clip,width=0.3\textwidth]{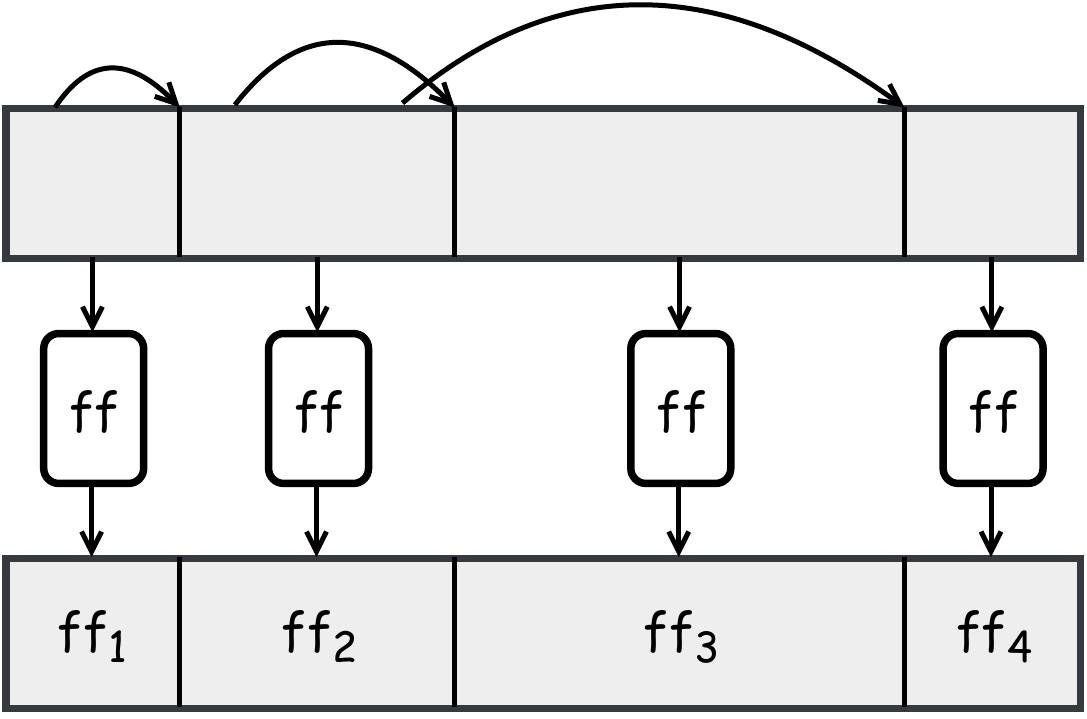}
	\caption{Example of the feature extraction process for a 3-FF Arbiter PUF~\cite{alkatheiri2017towards}.}
	\label{fig:ff-puf feature example}
\end{figure}

Based on Eq.~\ref{con:FF-PUF responses}, Alkatheiri \textit{et al.}~\cite{alkatheiri2017towards} provided a challenge transform mode of FF-APUF. Fig.~\ref{fig:ff-puf feature example} illustrates an example of feature extraction process for a 3-loop FF-APUF. The arrow above the challenge
points the starting and ending point of each feed-forward loop~\cite{alkatheiri2017towards}. That is, for FF-APUF with $k$ loops, the challenge is divided into $k+ 1$ sub challenges with each ending point serving as the division point. Then, sub challenges are transformed into feature vector respectively. The final feature vector is the concatenation of each individual feature vector corresponding to each sub challenge. It should be noted that for an $n$-bit challenge, the ($n-k$)-bit feature vector is obtained after transformation. 

\subsection{$Mn_{S_{1},S_{2},S_{3}}$ Arbiter-PUF}

\begin{figure}[h]
	\centering
	\includegraphics[trim=0 0 0 0,clip,width=0.50\textwidth]{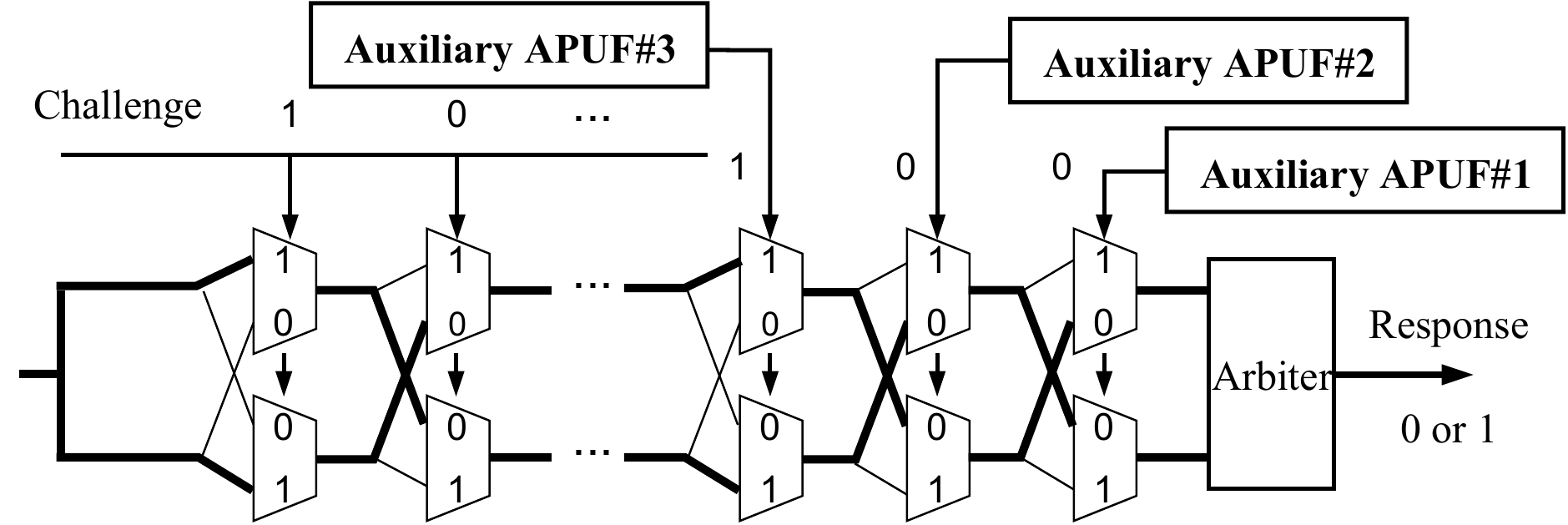}
	\caption{$Mn_{S_{1},S_{2},S_{3}}$-APUF~\cite{ebrahimabadi2021novel}}
	\label{fig:MnS1S2S3}
\end{figure}

Ebrahimabadi \textit{et al.}~\cite{ebrahimabadi2021novel} showed that the latter bits of the APUF challenge had a higher impact on determining the response (i.e., to be `1'/`0') than former challenge bits. Specifically, the Most Significant Bit (MSB), $c[n-1]$, is the most influential challenge bit on the response of an APUF. 

Based on the above observation, they proposed an APUF variant to increase its modeling resilience, termed as $Mn_{S_{1},S_{2},S_{3}}$-APUF, which consists of one main APUF and three auxiliary APUFs. The $n$ represents the stage of the main APUF, $S_1$, $S_2$ and $S_3$ denote the number of stages (i.e. size for simplicity) of the auxiliary APUFs that drive the first, second, and third most significant bit of the main APUF’s challenge, respectively. Note that when the size of auxiliary APUFs is less than $n$ bits, they are fed with a subset of the $n$-bit challenge of the main APUF~\cite{ebrahimabadi2021novel}. 

\subsection{$(x,y)$-$i$PUF}
\begin{figure}[h]
	\centering
	\includegraphics[trim=0 0 0 0,clip,width=0.50\textwidth]{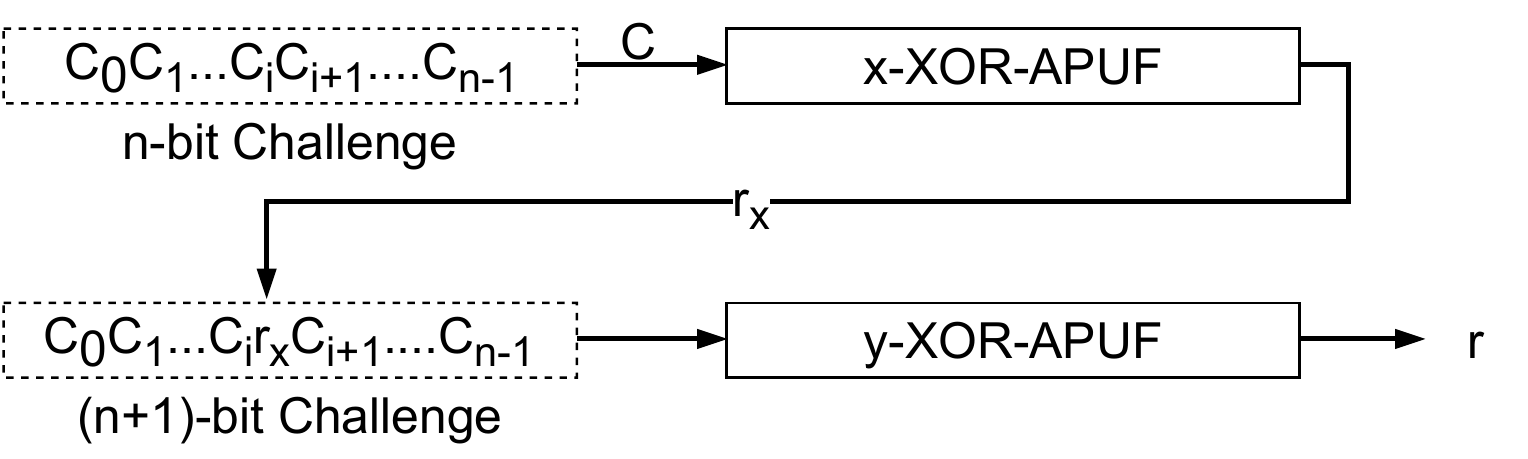}
	\caption{$n$-bit (x,y)-iPUF~\cite{nguyen2019interpose}.}
	\label{fig:ipuf}
\end{figure}

The interpose PUF (iPUF)~\cite{nguyen2019interpose} is also a challenge obfuscated PUF, which consists of two XOR-APUFs. As shown in Fig.~\ref{fig:ipuf}, the response of bottom $x$-XOR-APUF (with $n$ challenge bits) is interposed to the challenge utilized by upper $y$-XOR-APUF (with $n+1$ challenge bits)~\cite{santikellur2019deep}. When the response of $x$-XOR-APUF is inserted into the middle of the challenge of $y$-XOR-APUF, the $(x, y)$-$i$PUF exhibits highest modeling resilience, in particular, against classical machine learning modeling attacks---Logistic Regression (LR), reliability based attacks and cryptanalytic attacks.

Santikellur \textit{et al.}~\cite{santikellur2019deep} used a MLP with three hidden layers to attack the ($x,y$)-iPUF, which successfully attacked $(3,3)$-iPUF and $(4,4)$-iPUF. Wisiol \textit{et al.}~\cite{wisiol2020splitting} proposed a splitting attack on $(x,y)$-iPUF, where bottom and upper XOR-APUFs are attacked separately.

\subsection{XOR-FF-APUF}
XORing multiple responses from multiple PUFs can increase modeling resilience as XORing serves as a nonlinearity injection~\cite{gao2020physical}.
Avvaru \textit{et al.}\cite{avvaru2020homogeneous} used FF-APUFs as the underlying PUF components to construct XOR-FF-APUF. 
 
There are two variants of XOR-FF-APUFs: homogeneous XOR-FF-APUFs and heterogeneous XOR-FF-APUFs. The former uses the same design for the FF-APUFs involved in XOR operation, that is, the FF loops are located in the same stages for all components~\cite{avvaru2020homogeneous}. While the latter uses different loop designs for the FF-APUFs involved in XOR operation. Compared with XOR-APUF, XOR-FF-APUF has higher modeling resilience resistance.

\subsection{OAX-FF-APUF}
Yao \textit{et al.}~\cite{yao2021design} recently proposed OAX-PUF by using all three basic logic operations of OR, AND and XOR to processing responses from multiple PUFs---XOR-PUF is inclusive to OAX-PUF. With extensive empirical evaluations~\cite{yao2021design}, it has been shown the the modeling resilience of OAX-APUF is generally no less than XOR-APUF given same number of underlying APUFs used---most cases, XOR-APUF is much better. In addition, the OAX-PUF has advantage over XOR-PUF in terms of reliability. 

\begin{figure}[h]
	\centering
	\includegraphics[trim=0 0 0 0,clip,width=0.50\textwidth]{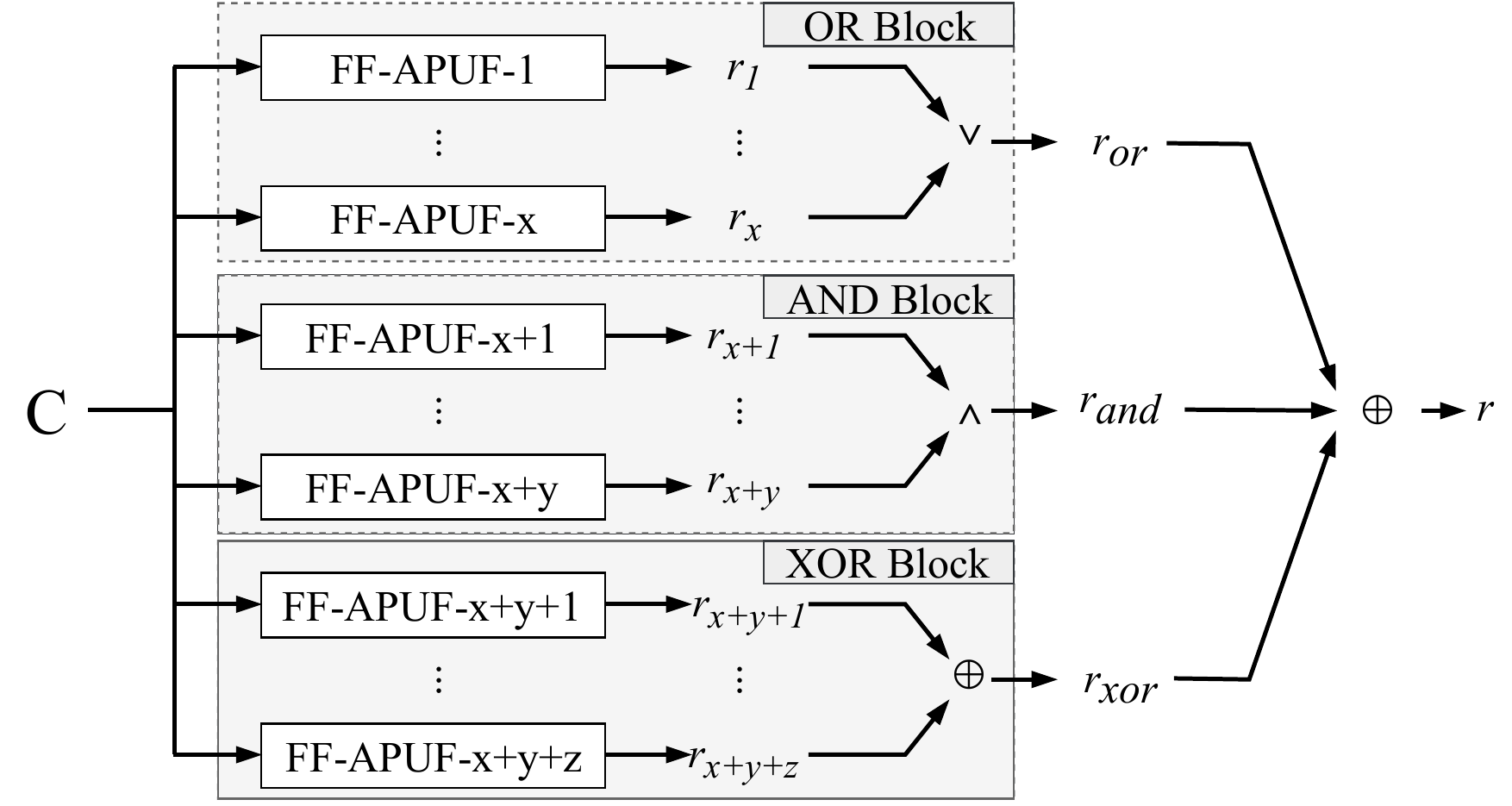}
	\caption{($x,y,z$)-OAX-FF-APUF.}
	\label{fig:oax-ff-puf}
\end{figure}

We propose to use FF-APUF as underlying PUF components to form OAX-FF-APUF. Similar to XOR-FF-APUF, there are two kinds of OAX-FF-APUFs: homogeneous OAX-FF-APUFs and heterogeneous OAX-FF-APUFs. The structure of OAX-FF-APUFs is shown in Fig.~\ref{fig:oax-ff-puf}, the responses of $x$ FF-APUFs are ORed to get $r_{or}$, the $y$ FF-APUFs’ responses are ANDed to get $r_{and}$, and the last responses of $z$ FF-APUFs are XORed to get $r_{xor}$. Finally, $r_{or}$, $r_{and}$, $r_{xor}$ are XORed to get $r$ of the ($x,y,z$)-OAX-FF-APUF. As the reliability of general OAX-PUF is provably higher than that of XOR-PUF~\cite{yao2021design}, the reliability of ($x,y,z$)-OAX-FF-APUF is  also higher than that of ($x+y+z$)-XOR-FF-APUF. Similary, it is expected that the modeling resilience of the ($x,y,z$)-OAX-FF-APUF is no less than ($x+y+z$)-XOR-FF-APUF.

\section{Deep Learning Techniques and Attack Mounting on CO-APUFs}
\label{sec:MLM}
We introduce the deep learning (DL) techniques including recurrent neural network, temporal convolutional network and multi-layer perception used in this work to evaluate the modeling resilience of the CO-APUFs. In addition, we describe the rationale behind the chosen of these DL techniques and how to mount each on attacking the CO-APUF and configure their hyper-parameters.

\subsection{Recurrent Neural Network (RNN)}
Recurrent neural networks (RNN) are dedicated sequential models that maintain a vector of hidden activations that are propagated through time~\cite{elman1990finding,werbos1990backpropagation,graves2012supervised,bai2018empirical}.
The intuitive appeal of recurrent modeling is that the hidden state can act as a representation of everything that has been seen so far in the sequence~\cite{bai2018empirical}. Because the basic RNN is difficult to train~\cite{bengio1994learning,pascanu2013difficulty,bai2018empirical}, more advanced architectures are used instead, such as long short-term memory (LSTM)~\cite{hochreiter1997long}, and gated recurrent unit (GRU)~\cite{cho2014properties}.

\vspace{0.2cm}
\noindent{\bf LSTM.} 

\begin{figure}[h]
  \centering
    \subfigure[LSTM Cell]{
     \label{fig:cell:a}
     \includegraphics[width=4.0cm]{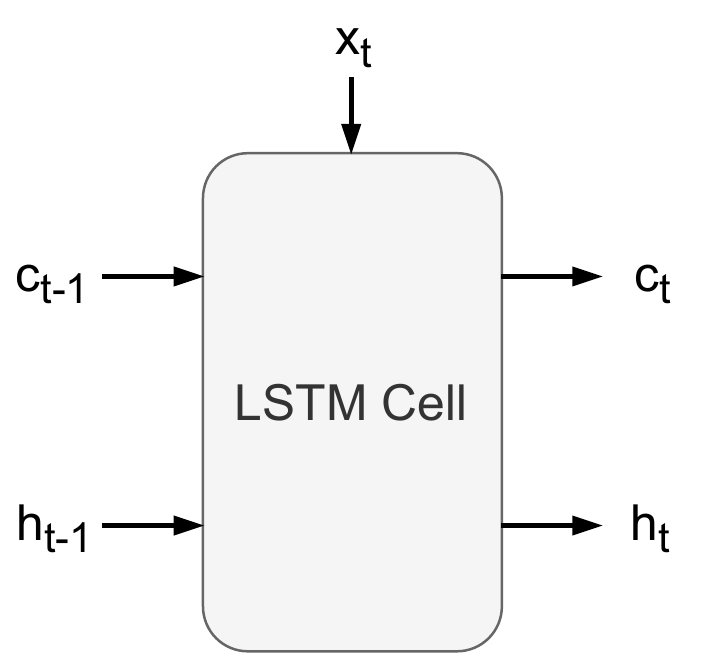}} 
    \subfigure[GRU Cell]{
     \label{fig:cell:b}
     \includegraphics[width=4.0cm]{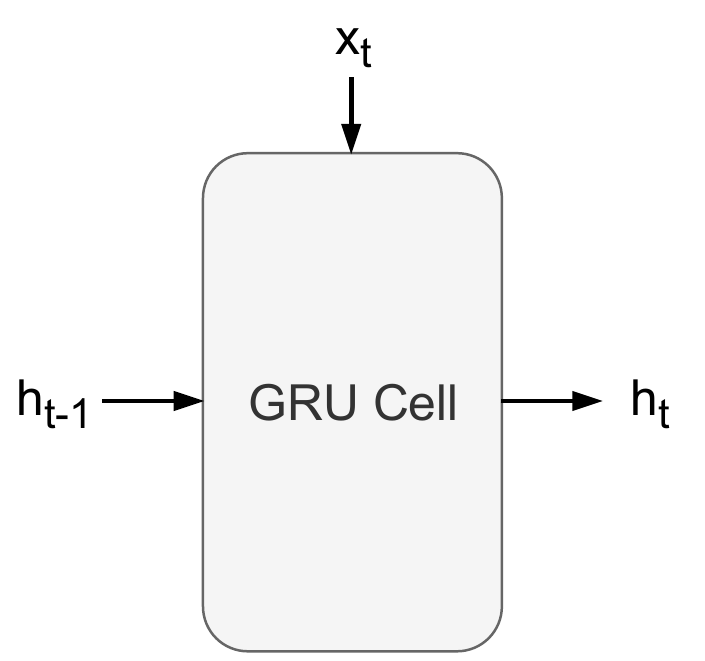}}
     \caption{LSTM and GRU cells.}
  \label{fig:cell}
 
\end{figure}

The simplified version of LSTM cell is shown in Fig.~\ref{fig:cell:a}. A LSTM cell is composed of three gates to control cell state and hidden state: forget gate $f_{t}$, input gate $i_{t}$ and output gate $o_{t}$. 

The LSTM cell works as below steps:
\begin{itemize}[]
\item
The forget gate selectively forgets the input from the previous cell.
That is after receiving the $c_{t-1}$ and $h_{t-1}$ of the previous node, current cell's forget gate decides which information to discard.
\item
The input gate is fed with current information $x_{t}$.
\item
Update cell state $c_{t}$.
\item
Output hidden state $h_{t}$.
\end{itemize}

\vspace{0.2cm}
\noindent{\bf GRU.} GRU is a variant of LSTM. Compared with LSTM, GRU combines the forget gate and the input gate into an update gate. GRU has two gates: reset gate $r_{t}$ and update gate $z_{t}$ (Eq.~\ref{eq:gru gates}). The GRU cell is shown in Fig.~\ref{fig:cell:b}.
\begin{equation}
\label{eq:gru gates}
    \begin{split}
        &r_{t}=\sigma(W_{r}\cdot [h_{t-1},x_{t}])\\ 
        &z_{t}=\sigma(W_{z}\cdot [h_{t-1},x_{t}]).
    \end{split}
\end{equation}
After receiving the $h_{t-1}$, GRU cell outputs the $h_{t}$ according to Eq.~\ref{eq:gru ht}.
\begin{equation}
\label{eq:gru ht}
    \begin{split}
        &\widetilde{h_{t}}=tanh(W\cdot [r_{t}*h_{t-1},x_{t}])\\ 
        &h_{t}=(1-z_{t})*h_{t-1}+z_{t}* \widetilde{h_{t}}.
    \end{split}
\end{equation}

\vspace{0.2cm}
\noindent{\bf Modeling CO-APUF: FF-APUF and XOR-FF-APUF.} Because each bit of the challenge influences on the response at different time points, the task of modeling APUF based on time delay, to a large extent, can be understood as the task of processing time series. In this case, the $n$-bit challenge can be understood as the time series input with $n$ time steps and one data point for each time step. Considering that the performance of GRU and LSTM is equal in many tasks, and GRU is easier to converge, we use GRU to attack FF-APUFs.

\begin{figure*}[h]
	\centering
	\includegraphics[trim=0 0 0 0,clip,width=0.80\textwidth]{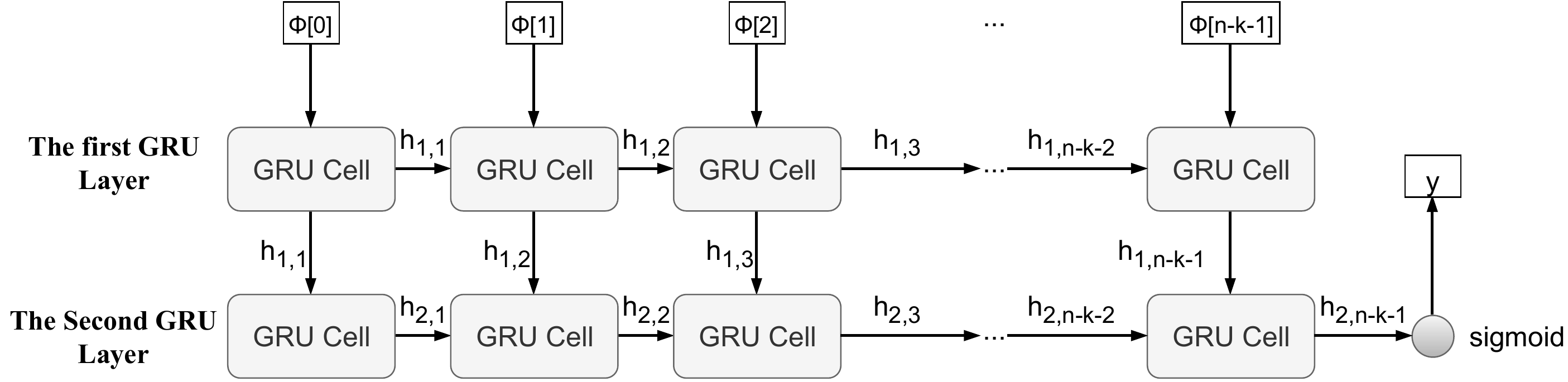}
	\caption{GRU structure attacking FF-APUF. The first GRU layer returns the outputs of all GRU cells in it. While the second GRU layer returns the output of the last GRU cell.}
	\label{fig:gru}
\end{figure*}

The GRU attack on FF-APUF is shown in Fig.~\ref{fig:gru}. The GRU structure has two GRU layers, and the input shape is $(batch\_size,n\_steps,n\_input)$, here $n\_input$ represents the number of inputs of each cell. When converting challenges to GRU input shapes, $n\_steps=n-k$, $n\_input=1$ ($n$ is the stage of FF-APUFs, $k$ represents the loops' number). The output of the first GRU layer is used as second GRU input layer, so it needs to return the outputs of all cells (the number of the cells is $n\_steps$). While the second GRU layer merely needs to return the output of last cell that represents the single response generated by a FF-APUF. 

The number of outputs of each cell needs careful analysis. The output shape of the first GRU layer is $(batch\_size,n-k,OUTPUT\_1)$. The output shape of the second GRU layer is $(batch\_size,OUTPUT\_2)$. Table \ref{tab:gru attack} summarizes three different settings/hyper-parameters for $OUTPUT\_1$ and $OUTPUT\_2$ we used. When the $OUTPUT\_1$ is 1, the attack efficacy is unsatisfactory, so we set $n-k$ (the same as the number of cells in GRU layer) according to our empirical trials. As for GRU\_1, setting the $OUTPUT\_2$ to 1 can be understood as obtaining a decisive value after $n-k$ GRU cells of the second GRU layer. As for GRU\_2, setting the OUTPUT\_2 to $2^{k+1}$ is inspired by the number of hidden layer neurons in MLP attack~\textit{et al.}~\cite{alkatheiri2017towards}. As for GRU\_3, it is a specific effective setting when attacking FF-APUF with 6 loops, because the first two settings take too long and the effect is unsatisfactory. Since GRU is easy to over fit, we add a drop out layer after the two GRU layers to avoid so. 
\begin{table}[]
\caption{GRU attack settings}

\label{tab:gru attack}
\centering
\begin{tabular}{|c|c|c|}
\hline
\textbf{GRU Attack Type} & \textbf{OUTPUT\_1} & \textbf{OUT\_PUT\_2}       \\ \hline
GRU\_1             & n-k                & 1                          \\ \hline
GRU\_2             & n-k                & $2^{k+1}$ \\ \hline
GRU\_3             & int((n-k)/2)       & $2^{k+2}$ \\ \hline
\end{tabular}
\end{table}

\subsection{Temporal Convolutional Network (TCN)}
Temporal convolutional network (TCN) is proposed by
S. Bai \textit{et al.}~\cite{bai2018empirical}, which can effectively deal with sequence modeling tasks, even better than other models. TCN combines best practices such as dilations and residual connections with the causal convolutions needed for autoregressive prediction~\cite{bai2018empirical}. 

\vspace{0.2cm}
\noindent{\bf Modeling CO-APUF: FF-APUF and XOR-FF-APUF.}  Since TCN have shown good performance in many tasks, we consider using it to attack FF-APUFs. The input shape of TCN is $(batch\_size,time\_steps,input\_dim)$, which is similar to the input shape of GRU.
We still need to transform the challenge vector matrix into three-dimensional, then we set $time\_steps=n-k$, $input\_dim=1$. The TCN attck is consisting of one TCN layer and one fully connection layer. The output shape of TCN layer is $(batch\_size,~nb\_filters)$. 
Here $nb\_filters$ is the number of filters used in the convolution layer. In the TCN attack, we consider $nb\_filters = n,n-k$, and $2 ^{k+ 1}$.
We found that $n-k$ or $n$ is with inferior performance compared with $2^{k+1}$ in our later experiments. 

\subsection{Multi-layer Perceptron (MLP)}
\begin{figure}[h]
	\centering
	\includegraphics[trim=0 0 0 0,clip,width=0.4\textwidth]{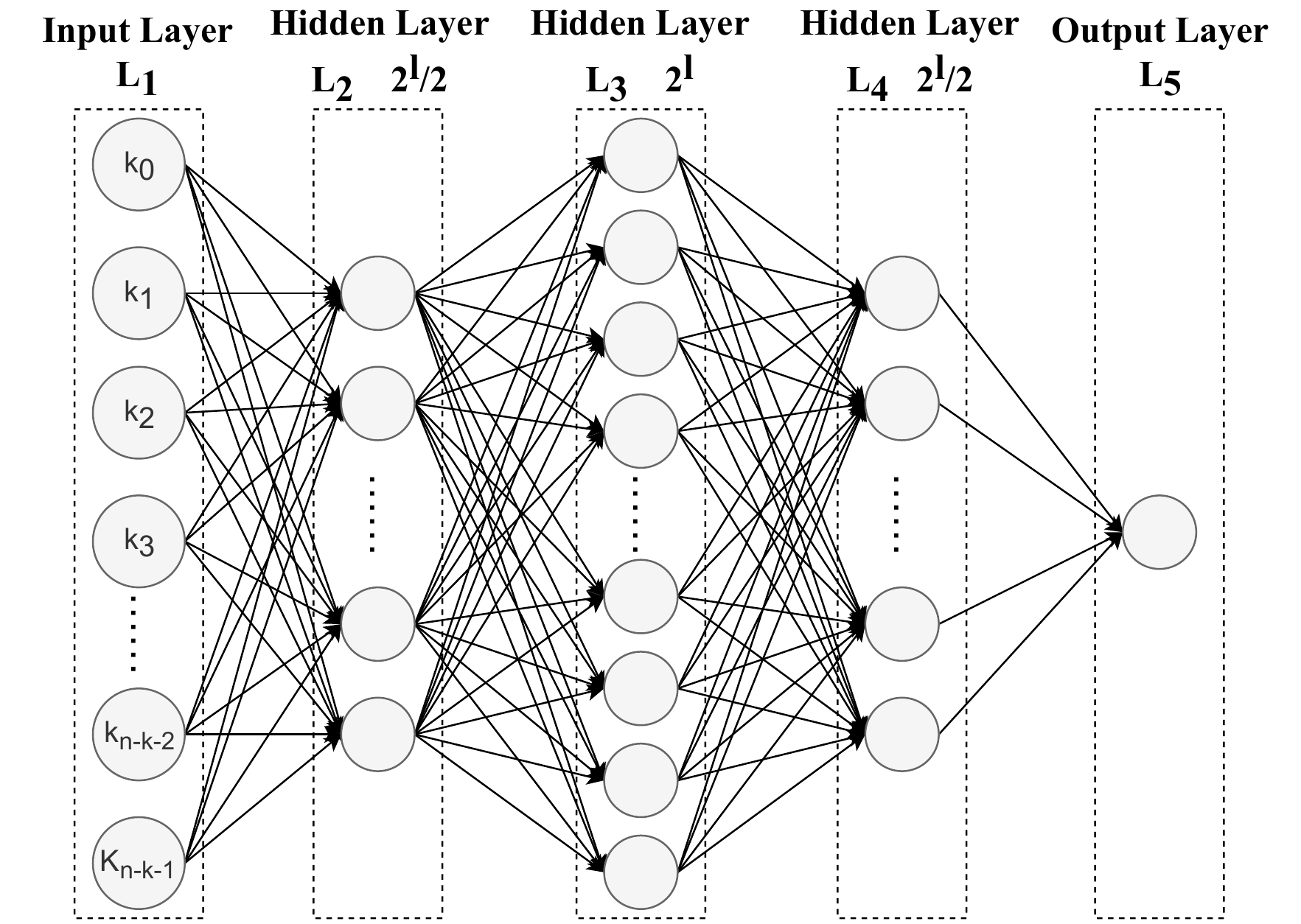}
	\caption{3-hidden layer MLP proposed by Mursi~\textit{et al.}~\cite{mursi2020fast} to attack \textit{XOR-APUFs}, here $l$ means the number of APUFs in XOR-APUF.}
	\label{fig:3-hidden layer MLP}
\end{figure}

A multi-layer perceptron (MLP)---a class of feed-forward artificial neural network (ANN)---has one input layer, one output layers and one or more hidden layers with many neurons stacked together~\cite{Pal1992Multilayer}. Fig.~\ref{fig:3-hidden layer MLP} shows the architecture of MLP with three layers. Except for neurons of the input layer, each neuron of MLP has an activation function that combines inputs and weights in a neuron, for instance the weighted sum, and imposes a threshold, such as \textsf{tanh}, \textsf{relu}, and \textsf{sigmoid}.

\vspace{0.2cm}
\noindent{\bf Modeling All CO-APUFs.} Mursi~\textit{et al.}~\cite{mursi2020fast} has used a MLP with three hidden layers to attack XOR-APUF and demonstrated high attacking accuracy. we consider using the MLP to attack other APUF variants, in particular, CO-APUFs. The MLP structure is shown in Fig.~\ref{fig:3-hidden layer MLP}, the activation function of the hidden layer is \textsf{tanh}, and the activation function of the output layer is \textsf{sigmoid}. The $l$ is tuned according to specific CO-APUFs:
\begin{itemize}
    \item For FF-APUF with $k$ loops, $l=k+1$, which can be understood as $k+1$ APUFs' responses decide the final response. 
    \item As for $M64_{32,16,8}$-APUF,  $l=4$, the reason is the same as FF-PUF. Specifically, there are one main APUF and three auxiliary APUFs---four in total. 
    \item For $z$-XOR-FF-APUF, $l= z+k+1$ or $l=z+k$.
    \item For ($x,y,z$)-OAX-FF-APUF, $l= x+y+z+k+1$ or $l=x+y+k-1$ or $l=x+y+z+k$.
    \item For ($x,y$)-iPUF, $l=\left \lceil \frac{x}{2}+y\right \rceil$, which is according to ~\cite{nguyen2019interpose} that ($x,y$)-iPUF's modeling resilience is akin to $(\frac{x}{2}+y)$-XOR-APUF.
\end{itemize}
Notably, for iPUF, in certain cases, $l$ needs to add 1 or subtract 1 to gain better attacking accuracy. 

\section{Reliability and Uniformity Evaluations}\label{sec:Exp}

We firstly describe our unified experimental settings, followed by comprehensive evaluations of each CO-APUFs in terms of two crucial PUF performance: reliability and uniformity. The modeling resilience evaluations are deferred to Section \ref{sec:Modeling Resilience}.

\begin{table}[]
\caption{BER and uniformity for $M64_{32,16,8}$-APUF and ($x,y$)-iPUFs}
\label{tab:PUFs}
\centering
\resizebox{0.5\textwidth}{!}{
\begin{tabular}{|c|c|c|c|}
\hline
\textbf{PUF}  & \textbf{\begin{tabular}[c]{@{}c@{}}Chall. Size\end{tabular}} & \textbf{BER} & \textbf{Uniformity$_{1}$} \\ \hline
$M64_{32,16,8}$-APUF & 64                                                                    & 0.223                    & 0.522                 \\ \hline
$(3,3)$-$i$PUF    & 64                                                                  & (0.118 ; 0.279)                                    & 0.502                 \\ \hline
$(4,4)$-$i$PUF    & 64                                                                    & (0.137 ; 0.323)                               & 0.508                 \\ \hline
$(5,5)$-$i$PUF    & 64                                                                    & (0.166 ; 0.362)                         & 0.509                 \\ \hline
$(1,7)$-$i$PUF    & 64                                                                    & (0.177 ; 0.389)                                & 0.486                 \\ \hline
\end{tabular}
}

    \begin{tablenotes}
      \footnotesize
      \item ($before; after$): $before$ and $after$ mean the BER is evaluated by setting $\sigma_{noise}=0.02$ and $\sigma_{noise}=0.05$, respectively.
    \end{tablenotes}
\end{table}

\begin{table}[]
\caption{FF-APUF's Loop Configuration (Config.)}
\label{tab:loopConfig}
\centering
\begin{tabular}{|c|c|c|}
\hline
\textbf{Loop Config. ID} & \textbf{Loop Nums} & \textbf{Start→End}   \\ \hline
Loop$_{A}$                   & 1                  & 15→25                \\ \hline
Loop$_{B}$                   & 2                  & 15→25,30             \\ \hline
Loop$_{C}$                   & 3                  & 15→25,30,35          \\ \hline
Loop$_{D}$                   & 3                  & 8→62;16→63;32→64     \\ \hline
Loop$_{E}$                   & 4                  & 15→25,30,35,40       \\ \hline
Loop$_{F}$                   & 5                  & 15→25,30,35,40,45    \\ \hline
Loop$_{G}$                   & 6                  & 15→25,30,35,40,45,50 \\ \hline
\end{tabular}
\end{table}

\begin{table}[]
\caption{BER and uniformity of FF-APUFs}
\label{tab:Feed Forward PUFs}
\centering
\begin{tabular}{|c|c|c|c|c|}
\hline
\textbf{\begin{tabular}[c]{@{}c@{}}Loop Config. \\ ID\end{tabular}} & \textbf{\begin{tabular}[c]{@{}c@{}}Loop \\ Nums \end{tabular}} & \textbf{\begin{tabular}[c]{@{}c@{}}Chall. Size \end{tabular}} & \textbf{BER}  & \textbf{Uniformity$_{1}$} \\ \hline
Loop$_{B}$                  & 2                  & 64                                                          & 0.071                        & 0.405                \\ \hline
Loop$_{C}$                  & 3                  & 64                                                       & 0.081                        & 0.399                \\ \hline
Loop$_{D}$                  & 3(*)                  & 64                                            & 0.201                       & 0.443                \\ \hline
Loop$_{E}$                  & 4                  & 64                                                     & 0.073                        & 0.400                \\ \hline
Loop$_{F}$                  & 5                  & 64                                                 & 0.080                        & 0.404                \\ \hline
Loop$_{G}$                  & 6                  & 64                                               & 0.075                       & 0.401                \\ \hline
\end{tabular}
\end{table}

\begin{table}[]
\caption{BER and uniformity of $z$-XOR-FF-APUFs}
\label{tab:XOR-FF-PUFs}
\centering
\resizebox{0.5\textwidth}{!}{
\begin{tabular}{|c|c|c|c|c|c|}
\hline
\textbf{\begin{tabular}[c]{@{}c@{}}Loop Config. \\ ID\end{tabular}}&
\textbf{\begin{tabular}[c]{@{}c@{}}Loop \\ Nums\end{tabular}} & \textbf{\begin{tabular}[c]{@{}c@{}}Chall. Size \end{tabular}}                                                   & \textbf{\begin{tabular}[c]{@{}c@{}}FF-APUF\\ Nums ($z$)\end{tabular}} & \textbf{BER}  & \textbf{Uniformity$_{1}$} \\ \hline
\multirow{5}{*}{Loop$_{A}$}                                           & \multirow{5}{*}{1}                                                      & \multirow{5}{*}{64}                                                                    & 2                                                              & 0.127              & 0.433       \\ \cline{4-6} 
                                                             &                                                                          &                                                                                            & 3                                                              & 0.225              & 0.507       \\ \cline{4-6} 
                                                             &                                                                          &                                                                                            & 4                                                              & 0.255              & 0.496       \\ \cline{4-6} 
                                                             &                                                                          &                                                                                            & 5                                                              & 0.357               & 0.494       \\ \cline{4-6} 
                                                             &                                                                          &                                                                                            & 6                                                              & 0.402              & 0.496       \\ \hline
\multirow{4}{*}{Loop$_{B}$}                                           & \multirow{4}{*}{2}                                                      & \multirow{4}{*}{64}                   & 2                                                              & 0.117               & 0.509       \\ \cline{4-6} 
                                                             &                                                                          &                                                                                            & 3                                                              & 0.212               & 0.503       \\ \cline{4-6} 
                                                             &                                                                          &                                                                                            & 4                                                              & 0.249               & 0.505       \\ \cline{4-6} 
                                                             &                                                                          &                                                                                            & 5                                                              & 0.321              & 0.499       \\ \hline
\multirow{3}{*}{Loop$_{C}$}                                           & \multirow{3}{*}{3}                                                      & \multirow{3}{*}{64}          & 2                                                              & 0.130               & 0.437       \\ \cline{4-6} 
                                                             &                                                                          &                                                                                            & 3                                                              & 0.246               & 0.506       \\ \cline{4-6} 
                                                             &                                                                          &                                                                                            & 4                                                              & 0.259              & 0.494       \\ \hline
\multirow{3}{*}{Loop$_{E}$}                                           & \multirow{3}{*}{4}                                                      & \multirow{3}{*}{64} & 2                                                              & 0.122               & 0.503       \\ \cline{4-6} 
                                                             &                                                                          &                                                                                            & 3                                                              & 0.218               & 0.496       \\ \cline{4-6} 
                                                             &                                                                          &                                                                                            & 4                                                              & 0.257               & 0.503       \\ \hline
\multirow{2}{*}{Loop$_{F}$}                                           & \multirow{2}{*}{5}                                                      & \multirow{2}{*}{64}    & 2                                                              & 0.135               & 0.429       \\ \cline{4-6} 
                                                             &                                                                          &                                                                                            & 3                                                              & 0.252             & 0.508       \\ \hline
Loop$_{G}$                                                            & 6                                                                      & 64                       & 2                                                              & 0.138               & 0.505       \\ \hline
\end{tabular}
}
\end{table}

\begin{table}[]
\caption{BER and uniformity of $(x,y,z)$-OAX-FF-APUFs }
\label{tab:OAX-FF-PUFs}
\centering
\resizebox{0.5\textwidth}{!}{
\begin{tabular}{|c|c|c|c|c|c|c|c|c|}
\hline
\textbf{\begin{tabular}[c]{@{}c@{}}Loop \\ Config.\\ ID \end{tabular}}     & \textbf{\begin{tabular}[c]{@{}c@{}}Loop \\ Nums \end{tabular}}  & \textbf{\begin{tabular}[c]{@{}c@{}}Chall. \\ Size\\ (bits)\end{tabular}} & \textbf{\begin{tabular}[c]{@{}c@{}}FF-APUF\\ Nums\\ ($x+y+z$)\end{tabular}} & \textbf{x} & \textbf{y} & \textbf{z} & \textbf{BER}  & \textbf{Uniformity$_{1}$} \\ \hline
\multirow{2}{*}{Loop$_{A}$} & \multirow{2}{*}{1} & \multirow{2}{*}{64}                                                      & \multirow{2}{*}{4}                                                          & 1          & 2          & 1          & 0.200                             & 0.523                     \\ \cline{5-9} 
                            &                    &                                                                          &                                                                             & 2          & 1          & 1          & 0.193                            & 0.562                     \\ \hline
\multirow{5}{*}{Loop$_{A}$} & \multirow{5}{*}{1} & \multirow{5}{*}{64}                                                      & \multirow{5}{*}{5}                                                          & 2          & 1          & 2          & 0.305                             & 0.492                     \\ \cline{5-9} 
                            &                    &                                                                          &                                                                             & 1          & 2          & 2          & 0.281                            & 0.497                     \\ \cline{5-9} 
                            &                    &                                                                          &                                                                             & 2          & 2          & 1          & 0.226                            & 0.488                     \\ \cline{5-9} 
                            &                    &                                                                          &                                                                             & 1          & 3          & 1          & 0.204                             & 0.440                     \\ \cline{5-9} 
                            &                    &                                                                          &                                                                             & 3          & 1          & 1          & 0.227                             & 0.540                     \\ \hline
\multirow{9}{*}{Loop$_{A}$} & \multirow{9}{*}{1} & \multirow{9}{*}{64}                                                      & \multirow{9}{*}{6}                                                          & 1          & 2          & 3          & 0.353                             & 0.494                     \\ \cline{5-9} 
                            &                    &                                                                          &                                                                             & 2          & 1          & 3          & 0.349                            & 0.497                     \\ \cline{5-9} 
                            &                    &                                                                          &                                                                             & 2          & 2          & 2          & 0.273                            & 0.507                     \\ \cline{5-9} 
                            &                    &                                                                          &                                                                             & 3          & 1          & 2          & 0.283                             & 0.507                     \\ \cline{5-9} 
                            &                    &                                                                          &                                                                             & 1          & 3          & 2          & 0.276                          & 0.484                     \\ \cline{5-9} 
                            &                    &                                                                          &                                                                             & 2          & 3          & 1          & 0.192                            & 0.495                     \\ \cline{5-9} 
                            &                    &                                                                          &                                                                             & 3          & 2          & 1          & 0.203                             & 0.512                     \\ \cline{5-9} 
                            &                    &                                                                          &                                                                             & 1          & 4          & 1          & 0.197                            & 0.468                     \\ \cline{5-9} 
                            &                    &                                                                          &                                                                             & 4          & 1          & 1          & 0.205                             & 0.519                     \\ \hline
\multirow{2}{*}{Loop$_{B}$} & \multirow{2}{*}{2} & \multirow{2}{*}{64}                                                      & \multirow{2}{*}{4}                                                          & 1          & 2          & 1          & 0.185                             & 0.529                     \\ \cline{5-9} 
                            &                    &                                                                          &                                                                             & 2          & 1          & 1          & 0.193                             & 0.534                     \\ \hline
\multirow{5}{*}{Loop$_{B}$} & \multirow{5}{*}{2} & \multirow{5}{*}{64}                                                      & \multirow{5}{*}{5}                                                          & 2          & 1          & 2          & 0.272                             & 0.498                     \\ \cline{5-9} 
                            &                    &                                                                          &                                                                             & 1          & 2          & 2          & 0.263                             & 0.495                     \\ \cline{5-9} 
                            &                    &                                                                          &                                                                             & 2          & 2          & 1          & 0.215                           & 0.503                     \\ \cline{5-9} 
                            &                    &                                                                          &                                                                             & 1          & 3          & 1          & 0.190                            & 0.466                     \\ \cline{5-9} 
                            &                    &                                                                          &                                                                             & 3          & 1          & 1          & 0.193                             & 0.551                     \\ \hline
\multirow{9}{*}{Loop$_{B}$} & \multirow{9}{*}{2} & \multirow{9}{*}{64}                                                      & \multirow{9}{*}{6}                                                          & 1          & 2          & 3          & 0.314                             & 0.501                     \\ \cline{5-9} 
                            &                    &                                                                          &                                                                             & 2          & 1          & 3          & 0.323                             & 0.499                     \\ \cline{5-9} 
                            &                    &                                                                          &                                                                             & 2          & 2          & 2          & 0.243                             & 0.510                     \\ \cline{5-9} 
                            &                    &                                                                          &                                                                             & 3          & 1          & 2          & 0.255                            & 0.504                     \\ \cline{5-9} 
                            &                    &                                                                          &                                                                             & 1          & 3          & 2          & 0.241                             & 0.505                     \\ \cline{5-9} 
                            &                    &                                                                          &                                                                             & 2          & 3          & 1          & 0.168                            & 0.567                     \\ \cline{5-9} 
                            &                    &                                                                          &                                                                             & 3          & 2          & 1          & 0.178                             & 0.538                     \\ \cline{5-9} 
                            &                    &                                                                          &                                                                             & 1          & 4          & 1          & 0.162                             & 0.565                     \\ \cline{5-9} 
                            &                    &                                                                          &                                                                             & 4          & 1          & 1          & 0.181                           & 0.515                     \\ \hline
\multirow{2}{*}{Loop$_{C}$} & \multirow{2}{*}{3} & \multirow{2}{*}{64}                                                      & \multirow{2}{*}{4}                                                          & 1          & 2          & 1          & 0.206                            & 0.512                     \\ \cline{5-9} 
                            &                    &                                                                          &                                                                             & 2          & 1          & 1          & 0.212                             & 0.558                     \\ \hline
\multirow{5}{*}{Loop$_{C}$} & \multirow{5}{*}{3} & \multirow{5}{*}{64}                                                      & \multirow{5}{*}{5}                                                          & 2          & 1          & 2          & 0.307                             & 0.499                     \\ \cline{5-9} 
                            &                    &                                                                          &                                                                             & 1          & 2          & 2          & 0.295                             & 0.498                     \\ \cline{5-9} 
                            &                    &                                                                          &                                                                             & 2          & 2          & 1          & 0.241                            & 0.496                     \\ \cline{5-9} 
                            &                    &                                                                          &                                                                             & 1          & 3          & 1          & 0.216                             & 0.457                     \\ \cline{5-9} 
                            &                    &                                                                          &                                                                             & 3          & 1          & 1          & 0.223                            & 0.542                     \\ \hline
\end{tabular}
}
\end{table}

\subsection{Setup}\label{sec:setup}
Following~\cite{ruhrmair2010modeling,becker2015pitfalls,nguyen2019interpose,wisiol2020splitting,sahoo2017multiplexer,yao2021design}, we use the standard means to simulate the APUF, which has been recognized as an efficient and common way when evaluating performance of APUF or its variants~\cite{ruhrmair2013puf,nguyen2019interpose}. We set the weights of APUF to follow Gaussian distribution $N (\mu, \sigma^{2})$, where $\mu = 0, \sigma = 1$. In order to simulate the unreliability of PUFs, we add random noise, which follows $N(\mu, \sigma_{noise}^{2})$. For for CO-APUFs, the $\sigma_{noise}$ is set to be 0.05. while for iPUF simulation, we have further considered $\sigma_{noise}=0.02$ to make sure that the iPUF bit error rate (BER) is under a reasonable range. When the $\sigma_{noise}$ is 0.05, the BER of APUF is 5.5\% $\sim $ 8.5\%. If $\sigma_{noise}$ is 0.02, the APUF's BER is about 2.12\% $\sim $ 3.17\%.

The diverse configurations of loops used in FF-APUFs, XOR-FF-APUFs and OAX-FF-APUFs are detailed in Table~\ref{tab:loopConfig}. The computing resource used to perform the DL attack is a common PC with an Intel(R) Core(TM) i5-6200U CPU, and a 12GB memory.

\subsection{Reliability and Uniformity}
The reliability measures the stability of the PUF under varying operating conditions (i.e. temperatures or/and voltages), which is usually measured by its complement metric of bit error rate (BER). Specifically, $\rm reliability = 1 - BER$. The closer BER to 0\%, the better. Uniformity indicates the bias of the PUF response. The closer to 0.5, the better. We use 10,000 challenges and repeatedly query them 11 times against the same PUF instance to evaluate the BER of simulated PUFs. Similarly, we use 10,000 challenges to count the ratio of `0' or `1' in their responses to evaluate the uniformity of the simulated PUFs.

Note that there are many specific configurations of each CO-APUFs (i.e. the number of loops and loop positions in FF-APUF), it is impractical to exhaustively evaluate all configurations. We instead select a number of configurations to present quantitative reports about the overall characteristic per CO-APUF.

\subsubsection{$Mn_{S_1,S_2,S_3}$-APUF and ($x,y$)-iPUFs} The $Mn_{S_1,S_2,S_3}$-APUF, to a large extent, shares similarities with the ($x,y$)-iPUFs. Because both of them utilize auxiliary APUFs to generate the obfuscated challenge bit(s). Here for the $Mn_{S_1,S_2,S_3}$-APUF, the main APUF has 64 stages, while the three auxiliary APUFs are with 32, 16, and 8 stages, respectively, following the same original setting in~\cite{ebrahimabadi2021novel}. For the ($x,y$)-iPUFs, several configurations are evaluated.

As for the uniformity, both of them exhibit good performance as the uniformity of each configuration is close to 0.5---with no notable deviation. 

As for the BER, both of them have somehow relatively notable degradation. Specifically, the $M64_{32,16,8}$-APUF is with a BER of 0.223. The ($x,y$)-iPUFs have higher BER when the $\sigma=0.05$. For instance, the ($5,5$)-iPUF and ($1,7$)-iPUF exhibit a BER of 0.362 and 0.389. This makes their practical usage to be challenging. Considering that the BER of underlying APUF can be reduced through e.g., customized ASIC (application-specific integrated circuit) design, it is feasible to reduce the underlying APUF BER to decrease the composite APUF's BER. Therefore, we have considered to lower the $\sigma=0.02$ for ($x,y$)-iPUFs, which significantly reduce the BER of the ($x,y$)-iPUFs. Note, the BER of the underlying APUF is about 2-3\%, which is realizable. Specifically, it was experimentally shown in~\cite{yu2016lockdown} that the BER of the ASIC APUF is about 4.5\% given the wide operating range ($-25\celsius$ to $85\celsius$) when the enrollment occurs at $25\celsius$. When the operating range is narrowed, this BER will be further reduced. As shown in~\cite{nguyen2019interpose}, the BER is between 0.66\% and 1.25\% when the APUF works under room temperature.

\subsubsection{FF-APUF Variants}
For the rest CO-APUFs including $z$-XOR-FF-APUF and ($x,y,z$)-OAX-FF-APUF, they are all FF-APUF variants.

As for the uniformity, the FF-APUF notably deviates from the 0.5, as shown in Table~\ref{tab:Feed Forward PUFs}. This aligns with other results~\cite{avvaru2020homogeneous}. As for the $z$-XOR-FF-APUF and ($x,y,z$)-OAX-FF-APUF, both can mitigate the worse uniformity, because the XOR operation has debiasing effect. As we can see from Table~\ref{tab:XOR-FF-PUFs}, when the $z$ increases in the $z$-XOR-FF-APUF, the uniformity improves to be close to 0.5.
Similarly, in Table~\ref{tab:OAX-FF-PUFs}, when the $z$ of ($x,y,z$)-OAX-FF-APUF increases, the uniformity improves. 

As for the BER, the FF-APUF degrades when the loop number increases, in particular, for loops with distinct (start, end) pairs. As shown in Table~\ref{tab:Feed Forward PUFs}, when loops share the same start point, the BER of the FF-APUF does not see obvious degradation even when the loop number increases (in particular, the end points increases), as for all rows except the fourth row (Loop$_{D}$) in Table~\ref{tab:Feed Forward PUFs}. In the fourth row, the (start, end) of each loop is different, which significantly deteriorates the BER. When the FF-APUF is given, the BER of the $z$-XOR-APUF and ($x,y,z$)-OAX-APUF increases given the increasing number of underlying FF-APUFs. But the BER of the ($x,y,z$)-OAX-APUF is much less than the $z$-XOR-APUF with same number of underlying FF-APUFs. For example, with six underlying FF-APUFs (Loop$_{A}$), the BER of ($2,3,1$)-OAX-FF-APUF is 0.192 while the $6$-XOR-FF-APUF is almost doubled to be 0.402. Therefore, the proposed ($x,y,z$)-OAX-APUF has obvious reliability advantage over the $z$-XOR-FF-APUF when enhancing the modeling resilience by compositing same number of underlying FF-APUFs.
\observ{In overall, the proposed ($x,y,z$)-OAX-APUF has the best reliability among these five CO-APUFs (i.e., 6 underlying APUF/FF-APUFs used). It also exhibits satisfactory uniformity performance}

\section{Modeling Resilience Evaluations}\label{sec:Modeling Resilience}
With same experimental setup and merely personal available computation resource (Section~\ref{sec:setup}), this part systematically assesses the modeling resilience of each CO-APUFs against DL attacks: diverse CO-APUF configurations and DL technique hyper-parameters are considered.

\subsection{FF-APUF}
\subsubsection{GRU} The GRU attacking performance is detailed in Table~\ref{tab: gru ff-apuf}. On the one hand, the GRU can successfully model the FF-APUF with two to six loops sharing the same start point (i.e. a single intermediate arbiter). On the other hand, the attacking complexity (i.e. larger number of training CRPs, and longer training time) increases as the loop number goes up. Specifically, the attacking accuracy is normally more than 92\% close to the FF-APUF reliability when the loop number is no more than 5. To attack the loop number of 6, only the GRU\_3 can successfully achieve close to 90\% accuracy with more than 6 hours training time. As for the three considered GRU configurations, in general, their empirical efficacy is GRU\_3 $>$ GRU\_2 $>$ GRU\_1.

\subsubsection{TCN} The TCN attacking performance against FF-APUF is detailed in Table~\ref{tab: tcn ff-puf}. The performance is similar to that of the GRU: attacking complexity increases as the loop number goes up. Overall, to achieve comparable attacking accuracy with the GRU, longer training times is required. As for the three considered $nb\_filters$ settings: $n$, $n-k$, and $2^{k+1}$, the $2^{k+1}$ is more preferable as it exhibits the best accuracy.

\subsubsection{MLP} We firstly reproduced MLP attack results according to the MLP configuration in~\cite{alkatheiri2017towards}\footnote{Note that \cite{alkatheiri2017towards} didn't specify their FF-APUFs' loop settings. So the reproductions are evaluated on the FF-APUFs' loop settings used in our work.}. The MLP consists of an input layer, a hidden layer and an output layer, in which the number of neurons in the hidden layer is $2 ^ {k + 1}$, and the hidden layer activation function adopts \textsf{tanh} when we reproduce (\cite{alkatheiri2017towards} did not specify what the hidden layer activation function adopts). The reproduced results are shown in Table~\ref{tab: mlp ff-puf re}, which are consistent with the results in~\cite{alkatheiri2017towards}. The MLP with three hidden layers can model the FF-APUFs with one intermediate arbiter and $k$ ($1\leq k \leq 6$) loops, with the accuracy more than 85\%. The MLP attacking performance on FF-APUF according to our MLP configuration is detailed in Table~\ref{tab: mlp ff-puf}. The attacking accuracy is always higher than that of~\cite{alkatheiri2017towards}, with about 3\% improvement. The attacking time is normally less than 20 minutes. It should be noted that the accuracy of the FF-APUF modeling attack has a certain relationship with its loop settings. 

\observ{All three DL techniques (i.e. GRU, TCN, and MLP) are capable of breaking the FF-APUF. Generally, the MLP is with the best efficacy (i.e. less than 20 min training time given same training CRPs)}

\observ{With same loop number, successfully attacking the FF-APUF that has different intermediate arbiter enabled loops ($loop_{C}$) has higher complexity than attacking that share same intermediate arbiter enabled loops ($loop_{D}$). But the former has notably decreased BER}

\begin{table*}
\caption{ GRU attacking performance against FF-APUFs}
\label{tab: gru ff-apuf}
\centering

\begin{tabular}{|cccccccccc|}
\hline
\multicolumn{1}{|c|}{\textbf{GRU Config.}} &
\multicolumn{1}{c|}{\textbf{Loop Config. ID}} &
\multicolumn{1}{c|}{\textbf{Loop Nums ($k$)}} & \multicolumn{1}{c|}{\textbf{\begin{tabular}[c]{@{}c@{}}Training\\ CRPs\end{tabular}}} & \multicolumn{1}{c|}{\textbf{\begin{tabular}[c]{@{}c@{}}Validation\\ CRPs\end{tabular}}} & \multicolumn{1}{c|}{\textbf{\begin{tabular}[c]{@{}c@{}}Test\\ CRPs\end{tabular}}} & \multicolumn{1}{c|}{\textbf{Epochs}} & \multicolumn{1}{c|}{\textbf{\begin{tabular}[c]{@{}c@{}}batch\\ size\end{tabular}}} & \multicolumn{1}{c|}{\textbf{\begin{tabular}[c]{@{}c@{}}Test\\ acc\end{tabular}}} &  \textbf{\begin{tabular}[c]{@{}c@{}}Training\\  Time\end{tabular}} \\ \hline

\multicolumn{1}{|c|}{GRU 1}           & \multicolumn{1}{c|}{Loop$_{B}$} 
& \multicolumn{1}{c|}{2}              & \multicolumn{1}{c|}{20,000}                                                            & \multicolumn{1}{c|}{5,000}                                                        & \multicolumn{1}{c|}{1,000}                                                         & \multicolumn{1}{c|}{200}             & \multicolumn{1}{c|}{200}                                                           & \multicolumn{1}{c|}{92.00\%}                                                  & 41 min                                                            \\ \hline
\multicolumn{1}{|c|}{GRU 2}           &
\multicolumn{1}{c|}{Loop$_{B}$}              &
\multicolumn{1}{c|}{2}              & \multicolumn{1}{c|}{20,000}                                                            & \multicolumn{1}{c|}{5,000}                                                        & \multicolumn{1}{c|}{1,000}                                                         & \multicolumn{1}{c|}{200}             & \multicolumn{1}{c|}{200}                                                           & \multicolumn{1}{c|}{92.40\%}                                                  & 42 min                                                            \\ \hline
\multicolumn{1}{|c|}{GRU 1}           &
\multicolumn{1}{c|}{Loop$_{C}$}              &
\multicolumn{1}{c|}{3}              & \multicolumn{1}{c|}{200,000}                                                           & \multicolumn{1}{c|}{50,000}                                                       & \multicolumn{1}{c|}{1,000}                                                         & \multicolumn{1}{c|}{40}              & \multicolumn{1}{c|}{200}                                                           & \multicolumn{1}{c|}{92.80\%}                                                  & 1.25h                                                             \\ \hline
\multicolumn{1}{|c|}{GRU 2}           &
\multicolumn{1}{c|}{Loop$_{C}$}              &
\multicolumn{1}{c|}{3}              & \multicolumn{1}{c|}{200,000}                                                           & \multicolumn{1}{c|}{50,000}                                                       & \multicolumn{1}{c|}{1,000}                                                         & \multicolumn{1}{c|}{40}              & \multicolumn{1}{c|}{200}                                                           & \multicolumn{1}{c|}{90.00\%}                                                  & 1.42 h                                                            \\ \hline
\multicolumn{1}{|c|}{GRU 1}           &
\multicolumn{1}{c|}{Loop$_{E}$}              &
\multicolumn{1}{c|}{4}              & \multicolumn{1}{c|}{200,000}                                                           & \multicolumn{1}{c|}{50,000}                                                       & \multicolumn{1}{c|}{1,000}                                                         & \multicolumn{1}{c|}{40}              & \multicolumn{1}{c|}{200}                                                           & \multicolumn{1}{c|}{91.90\%}                                                  & 1.22 h                                                            \\ \hline
\multicolumn{1}{|c|}{GRU 2}           &
\multicolumn{1}{c|}{Loop$_{E}$}              &
\multicolumn{1}{c|}{4}              & \multicolumn{1}{c|}{200,000}                                                           & \multicolumn{1}{c|}{50,000}                                                       & \multicolumn{1}{c|}{1,000}                                                         & \multicolumn{1}{c|}{40}              & \multicolumn{1}{c|}{200}                                                           & \multicolumn{1}{c|}{92.80\%}                                        & 1.62 h                                                            \\ \hline
\multicolumn{1}{|c|}{GRU 1}           &
\multicolumn{1}{c|}{Loop$_{F}$}              &
\multicolumn{1}{c|}{5}              & \multicolumn{1}{c|}{200,000}                                                           & \multicolumn{1}{c|}{50,000}                                                       & \multicolumn{1}{c|}{1,000}                                                         & \multicolumn{1}{c|}{40}              & \multicolumn{1}{c|}{200}                                                           & \multicolumn{1}{c|}{73.40\%}                                                   & 1.19 h                                                            \\ \hline
\multicolumn{1}{|c|}{GRU 2}           &
\multicolumn{1}{c|}{Loop$_{F}$}              &
\multicolumn{1}{c|}{5}              & \multicolumn{1}{c|}{200,000}                                                           & \multicolumn{1}{c|}{50,000}                                                       & \multicolumn{1}{c|}{1,000}                                                         & \multicolumn{1}{c|}{40}              & \multicolumn{1}{c|}{200}                                                           & \multicolumn{1}{c|}{93.00\%}                                                 & 2.3 h                                                             \\ \hline
\multicolumn{1}{|c|}{GRU 3 + Dropout}  &
\multicolumn{1}{c|}{Loop$_{G}$}              &
\multicolumn{1}{c|}{6}              & \multicolumn{1}{c|}{200,000}                                                           & \multicolumn{1}{c|}{50,000}                                                       & \multicolumn{1}{c|}{1,000}                                                         & \multicolumn{1}{c|}{40}              & \multicolumn{1}{c|}{200}                                                           & \multicolumn{1}{c|}{89.70\%}                                                   & 6.36 h                                                            \\ \hline
\end{tabular}
\end{table*}

\begin{table*}
\caption{ TCN attacking performance against FF-APUFs}
\label{tab: tcn ff-puf}
\centering
\begin{tabular}{|c|c|c|c|c|c|c|c|c|c|}
\hline
\textbf{Loop Config. ID}&
\textbf{Loop Nums ($k$)} & \textbf{\begin{tabular}[c]{@{}c@{}}Training \\ CRPs\end{tabular}} & \textbf{\begin{tabular}[c]{@{}c@{}}Validation \\ CRPs\end{tabular}} & \textbf{\begin{tabular}[c]{@{}c@{}}Test\\ CRPs\end{tabular}} & \textbf{Epochs} & \textbf{\begin{tabular}[c]{@{}c@{}}batch\\ size\end{tabular}} & \textbf{\begin{tabular}[c]{@{}c@{}}Test\\ acc\end{tabular}} & \textbf{\begin{tabular}[c]{@{}c@{}}Training\\ Time\end{tabular}} & \textbf{\begin{tabular}[c]{@{}c@{}}nb\_filters\\ config.\end{tabular}}        \\ \hline
Loop$_{B}$&2                                                                  & 20,000                                                             & 5,000                                                                & 1,000                                                         & 100             & 200                                                           & 91.50\%                                                     & 44 min                                                           & 62                         \\ \hline
Loop$_{B}$&2                                                                  & 20,000                                                             & 5,000                                                                & 1,000                                                         & 50              & 200                                                           & 89.30\%                                                      & 46 min                                                           & 64                         \\ \hline
Loop$_{C}$&3                                                                  & 20,000                                                             & 5,000                                                                & 1,000                                                         & 100             & 200                                                           & 87.40\%                                                     & 1.5 h                                                            & 61                         \\ \hline
Loop$_{E}$&4                                                                  & 200,000                                                            & 50,000                                                               & 1,000                                                         & 100             & 200                                                           & \textbf{93.00\%}                                            & 8.8 h                                                            & $2^{k+1}$ \\ \hline
Loop$_{F}$&5                                                                  & 200,000                                                            & 50,000                                                               & 1,000                                                         & 30              & 200                                                           & \textbf{90.50\%}                                            & 2.15 h                                                           & $2^{k}$   \\ \hline
Loop$_{F}$&5                                                                  & 200,000                                                            & 50,000                                                               & 1,000                                                         & 30              & 200                                                           & \textbf{91.70\%}                                            & 4.43 h                                                           & $2^{k+1}$ \\ \hline
Loop$_{G}$&6                                                                  & 200,000                                                            & 50,000                                                               & 1,000                                                         & 30              & 200                                                           & \textbf{89.30\%}                                             & 10.22 h                                                          & $2^{k+1}$ \\ \hline
\end{tabular}
\end{table*}

\begin{table*}
\caption{MLP attacking performance against FF-APUFs (reproduction according to MLP configuration in~\cite{alkatheiri2017towards})}
\label{tab: mlp ff-puf re}
\centering
\begin{tabular}{|ccccccccc|}
\hline
\multicolumn{1}{|c|}{\textbf{Loop Config. ID}}&
\multicolumn{1}{c|}{\textbf{Loop Nums ($k$)}} & \multicolumn{1}{c|}{\textbf{Training}} & \multicolumn{1}{c|}{\textbf{Validation}} & \multicolumn{1}{c|}{\textbf{Test}} & \multicolumn{1}{c|}{\textbf{Epochs}} & \multicolumn{1}{c|}
{\textbf{\begin{tabular}[c]{@{}c@{}}batch \\ size\end{tabular}}}& \multicolumn{1}{c|}{\textbf{\begin{tabular}[c]{@{}c@{}}Test \\ acc\end{tabular}}} &  \textbf{\begin{tabular}[c]{@{}c@{}}Training\\  Time\end{tabular}} \\ 
\hline

\multicolumn{1}{|c|}{Loop$_{B}$}&
\multicolumn{1}{c|}{2}             & \multicolumn{1}{c|}{20,000}             & \multicolumn{1}{c|}{5,000}                & \multicolumn{1}{c|}{1,000}          & \multicolumn{1}{c|}{100}             & \multicolumn{1}{c|}{20}                   & \multicolumn{1}{c|}{93.60\%}                                                             & 2 min                                                             \\ \hline
\multicolumn{1}{|c|}{Loop$_{C}$}&
\multicolumn{1}{c|}{3}             & \multicolumn{1}{c|}{20,000}             & \multicolumn{1}{c|}{5,000}                & \multicolumn{1}{c|}{1,000}          & \multicolumn{1}{c|}{100}             & \multicolumn{1}{c|}{20}                  & \multicolumn{1}{c|}{88.50\%}                                                      &  1.89 min                                                             \\ \hline
\multicolumn{1}{|c|}{Loop$_{E}$}&  
\multicolumn{1}{c|}{4}             & \multicolumn{1}{c|}{200,000}            & \multicolumn{1}{c|}{50,000}               & \multicolumn{1}{c|}{1,000}          & \multicolumn{1}{c|}{50}              & \multicolumn{1}{c|}{200}                  & \multicolumn{1}{c|}{92.40\%}                                                      &  2.4 min                                                           \\ \hline
\multicolumn{1}{|c|}{Loop$_{F}$} &
\multicolumn{1}{c|}{5}             & \multicolumn{1}{c|}{200,000}            & \multicolumn{1}{c|}{50,000}               & \multicolumn{1}{c|}{1,000}          & \multicolumn{1}{c|}{50}              & \multicolumn{1}{c|}{200}                  & \multicolumn{1}{c|}{89.40\%}                                                      &  2.51 min                                                          \\ \hline
\multicolumn{1}{|c|}{Loop$_{G}$} &
\multicolumn{1}{c|}{6}             & \multicolumn{1}{c|}{200,000}            & \multicolumn{1}{c|}{50,000}               & \multicolumn{1}{c|}{1,000}          & \multicolumn{1}{c|}{50}              & \multicolumn{1}{c|}{200}                  & \multicolumn{1}{c|}{86.10\%}                                                      &  2.75 min                                                          \\ \hline
\end{tabular}
\end{table*}

\begin{table*}
\caption{MLP attacking performance against FF-APUFs and $M64_{32,16,8}$ APUF (our MLP configuration as specified in the last four columns)
}
\label{tab: mlp ff-puf}
\centering
 \resizebox{\textwidth}{!}{
\begin{tabular}{|ccccccccccccc|}
\hline
\multicolumn{1}{|c|}{\textbf{Loop Config. ID}}   &\multicolumn{1}{c|}{\textbf{Loop Nums ($k$)}}   & \multicolumn{1}{c|}{\textbf{Training}} & \multicolumn{1}{c|}{\textbf{Validation}} & \multicolumn{1}{c|}{\textbf{Test}} & \multicolumn{1}{c|}{\textbf{Epochs}} & \multicolumn{1}{c|}{\textbf{\begin{tabular}[c]{@{}c@{}}batch \\size\end{tabular}}} & \multicolumn{1}{c|}{\textbf{\begin{tabular}[c]{@{}c@{}}Test \\ acc\end{tabular}}} & \multicolumn{1}{c|}{\textbf{\begin{tabular}[c]{@{}c@{}}Training\\  Time\end{tabular}}} & \multicolumn{1}{c|}{\textbf{layer1}} & \multicolumn{1}{c|}{\textbf{layer2}} & \multicolumn{1}{c|}{\textbf{layer3}} & \textbf{$l$} \\ \hline
\multicolumn{13}{|c|}{\textbf{Single Intermediate Arbiter}}                                                                                                                                                                                                                                                                                                                                                                                                                                                                                                                                                                                                                                                     \\ \hline
\multicolumn{1}{|c|}{Loop$_{B}$}&\multicolumn{1}{c|}{2}               & \multicolumn{1}{c|}{20,000}             & \multicolumn{1}{c|}{5,000}                & \multicolumn{1}{c|}{1,000}          & \multicolumn{1}{c|}{100}             & \multicolumn{1}{c|}{20}                   & \multicolumn{1}{c|}{95.50\%}                                                             & \multicolumn{1}{c|}{2.09 min}                                                          & \multicolumn{1}{c|}{4}               & \multicolumn{1}{c|}{8}               & \multicolumn{1}{c|}{4}               & 3          \\ \hline
\multicolumn{1}{|c|}{Loop$_{C}$}               & \multicolumn{1}{c|}{3}               & \multicolumn{1}{c|}{20,000}             & \multicolumn{1}{c|}{5,000}                & \multicolumn{1}{c|}{1,000}          & \multicolumn{1}{c|}{100}             & \multicolumn{1}{c|}{20}                   & \multicolumn{1}{c|}{91.30\%}                                                             & \multicolumn{1}{c|}{2.15 min}                                                          & \multicolumn{1}{c|}{8}               & \multicolumn{1}{c|}{16}              & \multicolumn{1}{c|}{8}               & 4          \\ \hline
\multicolumn{1}{|c|}{Loop$_{E}$}               &\multicolumn{1}{c|}{4}               & \multicolumn{1}{c|}{200,000}            & \multicolumn{1}{c|}{50,000}               & \multicolumn{1}{c|}{1,000}          & \multicolumn{1}{c|}{50}              & \multicolumn{1}{c|}{200}                  & \multicolumn{1}{c|}{92.10\%}                                                         & \multicolumn{1}{c|}{1.52 min}                                                          & \multicolumn{1}{c|}{16}              & \multicolumn{1}{c|}{32}              & \multicolumn{1}{c|}{16}              & 5          \\ \hline
\multicolumn{1}{|c|}{Loop$_{F}$}               &\multicolumn{1}{c|}{5}               & \multicolumn{1}{c|}{200,000}            & \multicolumn{1}{c|}{50,000}               & \multicolumn{1}{c|}{1,000}          & \multicolumn{1}{c|}{50}              & \multicolumn{1}{c|}{200}                  & \multicolumn{1}{c|}{92.80\%}                                                           & \multicolumn{1}{c|}{11.1 min}                                                          & \multicolumn{1}{c|}{32}              & \multicolumn{1}{c|}{64}              & \multicolumn{1}{c|}{32}              & 6          \\ \hline
\multicolumn{1}{|c|}{Loop$_{G}$}               &\multicolumn{1}{c|}{6}               & \multicolumn{1}{c|}{200,000}            & \multicolumn{1}{c|}{50,000}               & \multicolumn{1}{c|}{1,000}          & \multicolumn{1}{c|}{50}              & \multicolumn{1}{c|}{200}                  & \multicolumn{1}{c|}{89.30\%}                                                            & \multicolumn{1}{c|}{18.62 min}                                                         & \multicolumn{1}{c|}{64}              & \multicolumn{1}{c|}{128}             & \multicolumn{1}{c|}{64}              & 7          \\ \hline

\multicolumn{13}{|c|}{\textbf{Three Intermediate Arbiters and $M64_{32,16,8}$-APUF}}                                                                                                                                                                                                                                                                                                                                                                                                                                                                                                                                                           \\ \hline
\multicolumn{1}{|c|}{Loop$_{D}$}               &\multicolumn{1}{c|}{\textbf{3(*)}}      & \multicolumn{1}{c|}{200,000}            & \multicolumn{1}{c|}{50,000}               & \multicolumn{1}{c|}{1,000}          & \multicolumn{1}{c|}{100}             & \multicolumn{1}{c|}{20}                   & \multicolumn{1}{c|}{91.90\%}                                                  & \multicolumn{1}{c|}{21.3min}                                                           & \multicolumn{1}{c|}{8}               & \multicolumn{1}{c|}{16}              & \multicolumn{1}{c|}{8}               & 4          \\ \hline

\multicolumn{2}{|c|}{$M64_{32,16,8}$-APUF}
& \multicolumn{1}{c|}{200,000}            & \multicolumn{1}{c|}{50,000}               & \multicolumn{1}{c|}{1,000}          & \multicolumn{1}{c|}{100}             & \multicolumn{1}{c|}{20}                   & \multicolumn{1}{c|}{93.90\%}                                                                        & \multicolumn{1}{c|}{20.4 min}                                                                  & \multicolumn{1}{c|}{8}            & \multicolumn{1}{c|}{16}              & \multicolumn{1}{c|}{8}               & 4          \\
\hline
\end{tabular}
 }
\end{table*}

\subsection{$z$-XOR-FF-APUF}
\subsubsection{GRU and TCN} We have applied the GRU and TCN on attacking the $z$-XOR-FF-APUF, as results shown in Table~\ref{tab: gru-tcn XOR-ff-puf}. However, none of them can satisfactorily model the $z$-XOR-FF-APUF even for the smallest loop number of one and $z=2$, where the attacking accuracy is 63.00\% and 52.10\% for TCN and GRU, respectively. As for TCN, the $nb\_filters$ is $k$. As for GRU attack, GRU\_2 is used to train two FF-APUF models at the same time given $z=2$, and the outputs of the two models are connected to a fully connected layer. This indicates that the TCN and GRU are ineffective to model the nonlinear operations (i.e. XOR).

\subsubsection{MLP} The MLP attacking performance is detailed in Table~\ref{tab: mlp XOR-ff-puf}. When the number of loops or/and $z$ increase, the required number of training CRPs are increased. However, the MLP can always efficiently model the $z$-XOR-FF-APUF for $z\le 5$. To be precise, the attacking accuracy is normally higher than the BER of the given $z$-XOR-FF-APUF configuration. Though when $z=6$ ($loop_{A}$), the accuracy is alike guessing, it should be noted that the BER of this $6$-XOR-FF-APUF case is 40.2\%, may render impractical for deployment. Therefore, it is imperative to devise underling FF-APUF with enhanced reliability.

\observ{The GRU and TCN are ineffective to attack $z$-XOR-FF-APUF, while our MLP can effectively break it for $z\le 5$. Increasing the $z>6$ can be resilient to the MLP attack, but with a trade-off of severely degraded reliability}

\begin{table*}[]
\caption{GRU and TCN attacking performance against $z$-XOR-FF-APUFs.}
\label{tab: gru-tcn XOR-ff-puf}
\centering
\begin{tabular}{|ccccccccccc|}
\hline
\multicolumn{1}{|c|}{\textbf{Attack}} & \multicolumn{1}{c|}{\textbf{Loop Config. ID}} &\multicolumn{1}{c|}{\textbf{Loop Nums ($k$)}} & \multicolumn{1}{c|}{\textbf{\begin{tabular}[c]{@{}c@{}}FF-APUF\\ Nums ($z$)\end{tabular}}} & \multicolumn{1}{c|}{\textbf{\begin{tabular}[c]{@{}c@{}}Training\\ CRPs\end{tabular}}} & \multicolumn{1}{c|}{\textbf{\begin{tabular}[c]{@{}c@{}}Validation\\ CRPs\end{tabular}}} & \multicolumn{1}{c|}{\textbf{\begin{tabular}[c]{@{}c@{}}Test\\ CRPs\end{tabular}}} & \multicolumn{1}{c|}{\textbf{Epochs}} & \multicolumn{1}{c|}{\textbf{\begin{tabular}[c]{@{}c@{}}batch\\ size\end{tabular}}} & \multicolumn{1}{c|}{\textbf{\begin{tabular}[c]{@{}c@{}}Test \\ acc\end{tabular}}} & \textbf{\begin{tabular}[c]{@{}c@{}}Training\\  Time\end{tabular}} \\ \hline

\multicolumn{1}{|c|}{TCN}             & \multicolumn{1}{c|}{Loop$_{A}$}                                                            &\multicolumn{1}{c|}{1}                                                            & \multicolumn{1}{c|}{2}                                                              & \multicolumn{1}{c|}{20,000}                                                            & \multicolumn{1}{c|}{5,000}                                                        & \multicolumn{1}{c|}{1,000}                                                         & \multicolumn{1}{c|}{100}             & \multicolumn{1}{c|}{200}                                                           & \multicolumn{1}{c|}{63.00\%}                                                      & 1.95 h                                                            \\ \hline
\multicolumn{1}{|c|}{GRU\_2}             & \multicolumn{1}{c|}{Loop$_{A}$}  
& \multicolumn{1}{c|}{1}                                                            & \multicolumn{1}{c|}{2}                                                              & \multicolumn{1}{c|}{20,000}                                                            & \multicolumn{1}{c|}{5,000}                                                        & \multicolumn{1}{c|}{1,000}                                                         & \multicolumn{1}{c|}{100}             & \multicolumn{1}{c|}{200}                                                           & \multicolumn{1}{c|}{52.10\%}                                                      & 47.92 min                                                            \\ \hline
\end{tabular}
\end{table*}

\begin{table*}[]
\caption{MLP attacking performance against $z$-XOR-FF-APUFs (our MLP configuration as specified in last four columns).}
\label{tab: mlp XOR-ff-puf}
\centering
\resizebox{\textwidth}{!}{
\begin{tabular}{|cccccccccccccc|}
\hline
\multicolumn{1}{|l|}{\textbf{\begin{tabular}[c]{@{}c@{}}Loop\\ Config. ID\end{tabular}}}     & \multicolumn{1}{c|}{\textbf{\begin{tabular}[c]{@{}c@{}}Loop \\ Nums ($k$)\end{tabular}}} & \multicolumn{1}{c|}{\textbf{\begin{tabular}[c]{@{}c@{}}FF-PUF\\ Nums ($z$)\end{tabular}}} & \multicolumn{1}{c|}{\textbf{\begin{tabular}[c]{@{}c@{}}Training\\ CRPs\end{tabular}}} & \multicolumn{1}{c|}{\textbf{\begin{tabular}[c]{@{}c@{}}Validation\\ CRPs\end{tabular}}} & \multicolumn{1}{c|}{\textbf{\begin{tabular}[c]{@{}c@{}}Test\\ CRPs\end{tabular}}} & \multicolumn{1}{c|}{\textbf{Epochs}} & \multicolumn{1}{c|}{\textbf{\begin{tabular}[c]{@{}c@{}}batch\\ size\end{tabular}}} & \multicolumn{1}{c|}{\textbf{\begin{tabular}[c]{@{}c@{}}Test \\ acc\end{tabular}}} & \multicolumn{1}{c|}{\textbf{\begin{tabular}[c]{@{}c@{}}Training\\  Time\end{tabular}}} & \multicolumn{1}{c|}{\textbf{layer1}} & \multicolumn{1}{c|}{\textbf{layer2}} & \multicolumn{1}{c|}{\textbf{layer3}} & \textbf{$l$} \\ \hline

\multicolumn{1}{|c|}{\multirow{6}{*}{Loop$_{A}$}} & \multicolumn{1}{c|}{\multirow{6}{*}{1}}      & \multicolumn{1}{c|}{2}                                                              & \multicolumn{1}{c|}{20,000}                                                           & \multicolumn{1}{c|}{5,000}                                                       & \multicolumn{1}{c|}{1,000}                                                        & \multicolumn{1}{c|}{100}             & \multicolumn{1}{c|}{20}                                                            & \multicolumn{1}{c|}{93.70\%}                                                      & \multicolumn{1}{c|}{2 min}                                                             & \multicolumn{1}{c|}{8}               & \multicolumn{1}{c|}{16}              & \multicolumn{1}{c|}{8}               & 4            \\ \cline{3-14} 
\multicolumn{1}{|c|}{}                            & \multicolumn{1}{c|}{}                        & \multicolumn{1}{c|}{3}                                                              & \multicolumn{1}{c|}{40,000}                                                           & \multicolumn{1}{c|}{10,000}                                                      & \multicolumn{1}{c|}{1,000}                                                        & \multicolumn{1}{c|}{100}             & \multicolumn{1}{c|}{20}                                                            & \multicolumn{1}{c|}{91.90\%}                                                      & \multicolumn{1}{c|}{4.3 min}                                                           & \multicolumn{1}{c|}{16}              & \multicolumn{1}{c|}{32}              & \multicolumn{1}{c|}{16}              & 5            \\ \cline{3-14} 
\multicolumn{1}{|c|}{}                            & \multicolumn{1}{c|}{}                        & \multicolumn{1}{c|}{4}                                                              & \multicolumn{1}{c|}{100,000}                                                          & \multicolumn{1}{c|}{25,000}                                                      & \multicolumn{1}{c|}{1,000}                                                        & \multicolumn{1}{c|}{100}             & \multicolumn{1}{c|}{20}                                                            & \multicolumn{1}{c|}{93.50\%}                                                      & \multicolumn{1}{c|}{11.1min}                                                           & \multicolumn{1}{c|}{32}              & \multicolumn{1}{c|}{64}              & \multicolumn{1}{c|}{32}              & 6            \\ \cline{3-14} 
\multicolumn{1}{|c|}{}                            & \multicolumn{1}{c|}{}                        & \multicolumn{1}{c|}{5}                                                              & \multicolumn{1}{c|}{400,000}                                                          & \multicolumn{1}{c|}{100,000}                                                     & \multicolumn{1}{c|}{1,000}                                                        & \multicolumn{1}{c|}{100}             & \multicolumn{1}{c|}{200}                                                           & \multicolumn{1}{c|}{90.10\%}                                                      & \multicolumn{1}{c|}{11.1min}                                                           & \multicolumn{1}{c|}{64}              & \multicolumn{1}{c|}{128}             & \multicolumn{1}{c|}{64}              & 7            \\ \cline{3-14} 
\multicolumn{1}{|c|}{}                            & \multicolumn{1}{c|}{}                        & \multicolumn{1}{c|}{6}                                                              & \multicolumn{1}{c|}{500,000}                                                          & \multicolumn{1}{c|}{125,000}                                                     & \multicolumn{1}{c|}{1,000}                                                        & \multicolumn{1}{c|}{100}             & \multicolumn{1}{c|}{200}                                                           & \multicolumn{1}{c|}{\textbf{52.80\%}}                                                      & \multicolumn{1}{c|}{15 min}                                                            & \multicolumn{1}{c|}{128}             & \multicolumn{1}{c|}{256}             & \multicolumn{1}{c|}{128}             & \textbf{8}   \\ \cline{3-14} 
\multicolumn{1}{|c|}{}                            & \multicolumn{1}{c|}{}                        & \multicolumn{1}{c|}{6}                                                              & \multicolumn{1}{c|}{500,000}                                                          & \multicolumn{1}{c|}{125,000}                                                     & \multicolumn{1}{c|}{1,000}                                                        & \multicolumn{1}{c|}{100}             & \multicolumn{1}{c|}{200}                                                           & \multicolumn{1}{c|}{\textbf{50.70\%}}                                                      & \multicolumn{1}{c|}{11 min}                                                            & \multicolumn{1}{c|}{64}              & \multicolumn{1}{c|}{128}             & \multicolumn{1}{c|}{64}              & 7            \\ \hline
\multicolumn{1}{|c|}{\multirow{4}{*}{Loop$_{B}$}} & \multicolumn{1}{c|}{\multirow{4}{*}{2}}      & \multicolumn{1}{c|}{2}                                                              & \multicolumn{1}{c|}{100,000}                                                          & \multicolumn{1}{c|}{25,000}                                                      & \multicolumn{1}{c|}{1,000}                                                        & \multicolumn{1}{c|}{100}             & \multicolumn{1}{c|}{20}                                                            & \multicolumn{1}{c|}{93.20\%}                                                      & \multicolumn{1}{c|}{10.38 min}                                                         & \multicolumn{1}{c|}{16}              & \multicolumn{1}{c|}{32}              & \multicolumn{1}{c|}{16}              & 5            \\ \cline{3-14} 
\multicolumn{1}{|c|}{}                            & \multicolumn{1}{c|}{}                        & \multicolumn{1}{c|}{3}                                                              & \multicolumn{1}{c|}{200,000}                                                          & \multicolumn{1}{c|}{50,000}                                                      & \multicolumn{1}{c|}{1,000}                                                        & \multicolumn{1}{c|}{100}             & \multicolumn{1}{c|}{20}                                                            & \multicolumn{1}{c|}{88.50\%}                                                      & \multicolumn{1}{c|}{22 min}                                                            & \multicolumn{1}{c|}{32}              & \multicolumn{1}{c|}{64}              & \multicolumn{1}{c|}{32}              & 6            \\ \cline{3-14} 
\multicolumn{1}{|c|}{}                            & \multicolumn{1}{c|}{}                        & \multicolumn{1}{c|}{4}                                                              & \multicolumn{1}{c|}{400,000}                                                          & \multicolumn{1}{c|}{100,000}                                                     & \multicolumn{1}{c|}{1,000}                                                        & \multicolumn{1}{c|}{100}             & \multicolumn{1}{c|}{200}                                                           & \multicolumn{1}{c|}{88.60\%}                                                      & \multicolumn{1}{c|}{9 min}                                                             & \multicolumn{1}{c|}{64}              & \multicolumn{1}{c|}{128}             & \multicolumn{1}{c|}{64}              & 7            \\ \cline{3-14} 
\multicolumn{1}{|c|}{}                            & \multicolumn{1}{c|}{}                        & \multicolumn{1}{c|}{5}                                                              & \multicolumn{1}{c|}{500,000}                                                          & \multicolumn{1}{c|}{125,000}                                                     & \multicolumn{1}{c|}{1,000}                                                        & \multicolumn{1}{c|}{100}             & \multicolumn{1}{c|}{200}                                                           & \multicolumn{1}{c|}{85.40\%}                                                      & \multicolumn{1}{c|}{11 min}                                                            & \multicolumn{1}{c|}{64}              & \multicolumn{1}{c|}{128}             & \multicolumn{1}{c|}{64}              & 7            \\ \hline
\multicolumn{1}{|c|}{\multirow{4}{*}{Loop$_{C}$}} & \multicolumn{1}{c|}{\multirow{4}{*}{3}}      & \multicolumn{1}{c|}{2}                                                              & \multicolumn{1}{c|}{100,000}                                                          & \multicolumn{1}{c|}{25,000}                                                      & \multicolumn{1}{c|}{1,000}                                                        & \multicolumn{1}{c|}{100}             & \multicolumn{1}{c|}{20}                                                            & \multicolumn{1}{c|}{88.10\%}                                                      & \multicolumn{1}{c|}{10.78 min}                                                         & \multicolumn{1}{c|}{32}              & \multicolumn{1}{c|}{64}              & \multicolumn{1}{c|}{32}              & 6            \\ \cline{3-14} 
\multicolumn{1}{|c|}{}                            & \multicolumn{1}{c|}{}                        & \multicolumn{1}{c|}{3}                                                              & \multicolumn{1}{c|}{500,000}                                                          & \multicolumn{1}{c|}{125,000}                                                     & \multicolumn{1}{c|}{1,000}                                                        & \multicolumn{1}{c|}{100}             & \multicolumn{1}{c|}{200}                                                           & \multicolumn{1}{c|}{85.20\%}                                                      & \multicolumn{1}{c|}{11 min}                                                            & \multicolumn{1}{c|}{64}              & \multicolumn{1}{c|}{128}             & \multicolumn{1}{c|}{64}              & 7            \\ \cline{3-14} 
\multicolumn{1}{|c|}{}                            & \multicolumn{1}{c|}{}                        & \multicolumn{1}{c|}{4}                                                              & \multicolumn{1}{c|}{500,000}                                                          & \multicolumn{1}{c|}{100,000}                                                     & \multicolumn{1}{c|}{1,000}                                                        & \multicolumn{1}{c|}{100}             & \multicolumn{1}{c|}{200}                                                           & \multicolumn{1}{c|}{82.20\%}                                                      & \multicolumn{1}{c|}{11.2 min}                                                          & \multicolumn{1}{c|}{64}              & \multicolumn{1}{c|}{128}             & \multicolumn{1}{c|}{64}              & \textbf{7}   \\ \cline{3-14} 
\multicolumn{1}{|c|}{}                            & \multicolumn{1}{c|}{}                        & \multicolumn{1}{c|}{4}                                                              & \multicolumn{1}{c|}{500,000}                                                          & \multicolumn{1}{c|}{125,000}                                                     & \multicolumn{1}{c|}{1,000}                                                        & \multicolumn{1}{c|}{100}             & \multicolumn{1}{c|}{200}                                                           & \multicolumn{1}{c|}{79.40\%}                                                      & \multicolumn{1}{c|}{15.2 min}                                                          & \multicolumn{1}{c|}{128}             & \multicolumn{1}{c|}{256}             & \multicolumn{1}{c|}{128}             & 8            \\ \hline
\multicolumn{1}{|c|}{\multirow{3}{*}{Loop$_{E}$}} & \multicolumn{1}{c|}{\multirow{3}{*}{4}}      & \multicolumn{1}{c|}{2}                                                              & \multicolumn{1}{c|}{500,000}                                                          & \multicolumn{1}{c|}{125,000}                                                     & \multicolumn{1}{c|}{1,000}                                                        & \multicolumn{1}{c|}{100}             & \multicolumn{1}{c|}{200}                                                           & \multicolumn{1}{c|}{87.20\%}                                                      & \multicolumn{1}{c|}{11 min}                                                            & \multicolumn{1}{c|}{64}              & \multicolumn{1}{c|}{128}             & \multicolumn{1}{c|}{64}              & 7            \\ \cline{3-14} 
\multicolumn{1}{|c|}{}                            & \multicolumn{1}{c|}{}                        & \multicolumn{1}{c|}{3}                                                              & \multicolumn{1}{c|}{500,000}                                                          & \multicolumn{1}{c|}{125,000}                                                     & \multicolumn{1}{c|}{1,000}                                                        & \multicolumn{1}{c|}{100}             & \multicolumn{1}{c|}{200}                                                           & \multicolumn{1}{c|}{83.40\%}                                                      & \multicolumn{1}{c|}{10.97 min}                                                         & \multicolumn{1}{c|}{64}              & \multicolumn{1}{c|}{128}             & \multicolumn{1}{c|}{64}              & 7            \\ \cline{3-14} 
\multicolumn{1}{|c|}{}                            & \multicolumn{1}{c|}{}                        & \multicolumn{1}{c|}{4}                                                              & \multicolumn{1}{c|}{500,000}                                                          & \multicolumn{1}{c|}{125,000}                                                     & \multicolumn{1}{c|}{1,000}                                                        & \multicolumn{1}{c|}{100}             & \multicolumn{1}{c|}{200}                                                           & \multicolumn{1}{c|}{78.90\%}                                                      & \multicolumn{1}{c|}{11 min}                                                            & \multicolumn{1}{c|}{64}              & \multicolumn{1}{c|}{128}             & \multicolumn{1}{c|}{64}              & 7            \\ \hline
\multicolumn{1}{|c|}{\multirow{2}{*}{Loop$_{F}$}} & \multicolumn{1}{c|}{\multirow{2}{*}{5}}      & \multicolumn{1}{c|}{2}                                                              & \multicolumn{1}{c|}{500,000}                                                          & \multicolumn{1}{c|}{125,000}                                                     & \multicolumn{1}{c|}{1,000}                                                        & \multicolumn{1}{c|}{100}             & \multicolumn{1}{c|}{200}                                                           & \multicolumn{1}{c|}{83.90\%}                                                      & \multicolumn{1}{c|}{11.2 min}                                                          & \multicolumn{1}{c|}{64}              & \multicolumn{1}{c|}{128}             & \multicolumn{1}{c|}{64}              & 7            \\ \cline{3-14} 
\multicolumn{1}{|c|}{}                            & \multicolumn{1}{c|}{}                        & \multicolumn{1}{c|}{3}                                                              & \multicolumn{1}{c|}{500,000}                                                          & \multicolumn{1}{c|}{125,000}                                                     & \multicolumn{1}{c|}{1,000}                                                        & \multicolumn{1}{c|}{100}             & \multicolumn{1}{c|}{200}                                                           & \multicolumn{1}{c|}{80.30\%}                                                      & \multicolumn{1}{c|}{10.9 min}                                                          & \multicolumn{1}{c|}{64}              & \multicolumn{1}{c|}{128}             & \multicolumn{1}{c|}{64}              & 7            \\ \hline
\multicolumn{1}{|c|}{\multirow{2}{*}{Loop$_{G}$}} & \multicolumn{1}{c|}{\multirow{2}{*}{6}}      & \multicolumn{1}{c|}{2}                                                              & \multicolumn{1}{c|}{500,000}                                                          & \multicolumn{1}{c|}{125,000}                                                     & \multicolumn{1}{c|}{1,000}                                                        & \multicolumn{1}{c|}{100}             & \multicolumn{1}{c|}{200}                                                           & \multicolumn{1}{c|}{78.40\%}                                                      & \multicolumn{1}{c|}{15.1 min}                                                          & \multicolumn{1}{c|}{128}             & \multicolumn{1}{c|}{256}             & \multicolumn{1}{c|}{128}             & 8            \\ \cline{3-14} 
\multicolumn{1}{|c|}{}                            & \multicolumn{1}{c|}{}                        & \multicolumn{1}{c|}{2}                                                              & \multicolumn{1}{c|}{500,000}                                                          & \multicolumn{1}{c|}{125,000}                                                     & \multicolumn{1}{c|}{1,000}                                                        & \multicolumn{1}{c|}{100}             & \multicolumn{1}{c|}{200}                                                           & \multicolumn{1}{c|}{81.50\%}                                                      & \multicolumn{1}{c|}{11 min}                                                            & \multicolumn{1}{c|}{64}              & \multicolumn{1}{c|}{128}             & \multicolumn{1}{c|}{64}              & \textbf{7}   \\ \hline
\end{tabular}
}
\end{table*}

\subsection{($x,y,z$)-OAX-FF-APUF}
As the GRU and TCN are incapable of modeling XOR-FF-APUF, we do not further consider them for attacking ($x,y,z$)-OAX-FF-APUF. Note that OAX-FF-APUF is inclusive of XOR-FF-APUF. The MLP attacking performance against ($x,y,z$)-OAX-FF-APUF is detailed in Table~\ref{tab: mlp oax-ff-puf}. We consider the loop number 1, 2 and 3. If the loop number is 1, the MLP can successfully break $(x,y,z)$-OAX-FF-APUF with accuracy more than 90\%. For underlying FF-APUFs with 2 loops, the MLP can model the $(x,y,z)$-OAX-FF-APUF with more than 85\% accuracy when $x+y+z\leq 5$. The modeling resilience will be further improved by increasing the loop number. As when the underlying FF-APUFs with 3 loops, none of the accuracy reaches 90\% or more. As for the selection of the number of hidden layer neurons ($l$), for the case of one loop, most MLP configurations can obtain better accuracy by using $l = x + y + z + k + 1 $. When the number of loops is greater than 1, $l = x + y + z + k $, or $l = x + y+z+k-1 $, is more appropriate, such as for ($1,2,2 $)-OAX-FF-APUF with 2 loops.

\observ{
The OAX-FF-APUF of up to $x+y+z=6$ is also breakable with out MLP hyper-parameter settings, although the BER of OAX-FF-APUF is greatly lower compared to that of XOR-FF-APUF with same underlying FF-APUFs. Since the attacking accuracy is closer or higher than the ($x, y, z$)-OAX-FF-APUF BER
}

\begin{table*}[]
\caption{MLP attacking performance against ($x,y,z$)-OAX-FF-APUFs (our MLP configuration as specified in last four columns)}
\label{tab: mlp oax-ff-puf}
\centering
\resizebox{\textwidth}{!}{
\begin{tabular}{|c|c|c|c|c|c|c|c|c|c|c|c|c|c|c|c|c|}
\hline
\textbf{\begin{tabular}[c]{@{}c@{}}Loop Config.\\ ID\end{tabular}} & \textbf{\begin{tabular}[c]{@{}c@{}}Loop\\ Nums\\ ($k$)\end{tabular}} & \textbf{\begin{tabular}[c]{@{}c@{}}FF-PUF\\ Nums\\ ($x+y+z$)\end{tabular}} & \textbf{x} & \textbf{y} & \textbf{z} & \textbf{\begin{tabular}[c]{@{}c@{}}Training\\ CRPs\end{tabular}} & \textbf{\begin{tabular}[c]{@{}c@{}}Validation\\ CRPs\end{tabular}} & \textbf{\begin{tabular}[c]{@{}c@{}}Test\\ CRPs\end{tabular}} & \textbf{Epochs} & \textbf{\begin{tabular}[c]{@{}c@{}}batch\\ size\end{tabular}} & \textbf{\begin{tabular}[c]{@{}c@{}}Test \\ acc\end{tabular}} & \textbf{\begin{tabular}[c]{@{}c@{}}Training\\  Time\end{tabular}} & \textbf{layer1} & \textbf{layer2} & \textbf{layer3} & \textbf{$l$} \\ \hline
\multirow{4}{*}{Loop$_{A}$}                                        & \multirow{4}{*}{1}                                                    & \multirow{4}{*}{4}                                                         & 1          & 2          & 1          & 100,000                                                          & 25,000                                                      & 1,000                                                        & 100             & 20                                                            & 94.80\%                                                      & 11.41 min                                                         & 16              & 32              & 16              & \textbf{5}   \\ \cline{4-17} 
                                                                   &                                                                       &                                                                            & 1          & 2          & 1          & 100,000                                                          & 25,000                                                      & 1,000                                                        & 100             & 20                                                            & 93.80\%                                                      & 12.1 min                                                          & 32              & 64              & 32              & 6            \\ \cline{4-17} 
                                                                   &                                                                       &                                                                            & 2          & 1          & 1          & 100,000                                                          & 25,000                                                      & 1,000                                                        & 100             & 20                                                            & 94.90\%                                                      & 11.98 min                                                         & 8               & 32              & 8               & 5            \\ \cline{4-17} 
                                                                   &                                                                       &                                                                            & 2          & 1          & 1          & 100,000                                                          & 25,000                                                      & 1,000                                                        & 100             & 20                                                            & 95.10\%                                                      & 11.37min                                                          & 32              & 64              & 32              & \textbf{6}   \\ \hline
\multirow{5}{*}{Loop$_{A}$}                                        & \multirow{5}{*}{1}                                                    & \multirow{5}{*}{5}                                                         & 2          & 1          & 2          & 500,000                                                          & 125,000                                                     & 1,000                                                        & 50              & 200                                                           & 91.40\%                                                      & 5.82 min                                                          & 64              & 128             & 64              & 7            \\ \cline{4-17} 
                                                                   &                                                                       &                                                                            & 1          & 2          & 2          & 500,000                                                          & 125,000                                                     & 1,000                                                        & 50              & 200                                                           & 91.20\%                                                      & 5.72 min                                                          & 64              & 128             & 64              & 7            \\ \cline{4-17} 
                                                                   &                                                                       &                                                                            & 2          & 2          & 1          & 500,000                                                          & 125,000                                                     & 1,000                                                        & 50              & 200                                                           & 93.80\%                                                      & 5.72 min                                                          & 64              & 128             & 64              & 7            \\ \cline{4-17} 
                                                                   &                                                                       &                                                                            & 1          & 3          & 1          & 500,000                                                          & 125,000                                                     & 1,000                                                        & 50              & 200                                                           & 94.10\%                                                      & 5.63 min                                                          & 64              & 128             & 64              & 7            \\ \cline{4-17} 
                                                                   &                                                                       &                                                                            & 3          & 1          & 1          & 500,000                                                          & 125,000                                                     & 1,000                                                        & 50              & 200                                                           & 93.80\%                                                      & 5.89 min                                                          & 64              & 128             & 64              & 7            \\ \hline
\multirow{9}{*}{Loop$_{A}$}                                        & \multirow{9}{*}{1}                                                    & \multirow{9}{*}{6}                                                         & 1          & 2          & 3          & 500,000                                                          & 125,000                                                     & 1,000                                                        & 50              & 200                                                           & 89.30\%                                                      & 7.99 min                                                          & 128             & 256             & 128             & 8            \\ \cline{4-17} 
                                                                   &                                                                       &                                                                            & 2          & 1          & 3          & 500,000                                                          & 125,000                                                     & 1,000                                                        & 50              & 200                                                           & 87.60\%                                                      & 8.1 min                                                           & 128             & 256             & 128             & 8            \\ \cline{4-17} 
                                                                   &                                                                       &                                                                            & 2          & 2          & 2          & 500,000                                                          & 125,000                                                     & 1,000                                                        & 50              & 200                                                           & 90.40\%                                                      & 7.61 min                                                          & 128             & 256             & 128             & 8            \\ \cline{4-17} 
                                                                   &                                                                       &                                                                            & 3          & 1          & 2          & 500,000                                                          & 125,000                                                     & 1,000                                                        & 50              & 200                                                           & 93.00\%                                                      & 7.42 min                                                          & 128             & 256             & 128             & 8            \\ \cline{4-17} 
                                                                   &                                                                       &                                                                            & 1          & 3          & 2          & 500,000                                                          & 125,000                                                     & 1,000                                                        & 50              & 200                                                           & 90.20\%                                                      & 7.4 min                                                           & 128             & 256             & 128             & 8            \\ \cline{4-17} 
                                                                   &                                                                       &                                                                            & 2          & 3          & 1          & 500,000                                                          & 125,000                                                     & 1,000                                                        & 50              & 200                                                           & 95.00\%                                                      & 7.5 min                                                           & 128             & 256             & 128             & 8            \\ \cline{4-17} 
                                                                   &                                                                       &                                                                            & 3          & 2          & 1          & 500,000                                                          & 125,000                                                     & 1,000                                                        & 50              & 200                                                           & 95.30\%                                                      & 7.4 min                                                           & 128             & 256             & 128             & 8            \\ \cline{4-17} 
                                                                   &                                                                       &                                                                            & 1          & 4          & 1          & 500,000                                                          & 125,000                                                     & 1,000                                                        & 50              & 200                                                           & 93.80\%                                                      & 7.43 min                                                          & 128             & 256             & 128             & 8            \\ \cline{4-17} 
                                                                   &                                                                       &                                                                            & 4          & 1          & 1          & 500,000                                                          & 125,000                                                     & 1,000                                                        & 50              & 200                                                           & 94.70\%                                                      & 7.33 min                                                          & 128             & 256             & 128             & 8            \\ \hline
\multirow{2}{*}{Loop$_{B}$}                                        & \multirow{2}{*}{2}                                                    & \multirow{2}{*}{4}                                                         & 1          & 2          & 1          & 400,000                                                          & 100,000                                                     & 1,000                                                        & 50              & 200                                                           & 91.90\%                                                      & 4.55 min                                                          & 64              & 128             & 64              & 7            \\ \cline{4-17} 
                                                                   &                                                                       &                                                                            & 2          & 1          & 1          & 400,000                                                          & 100,000                                                     & 1,000                                                        & 50              & 200                                                           & 92.00\%                                                      & 4.76 min                                                          & 64              & 128             & 64              & 7            \\ \hline
\multirow{10}{*}{Loop$_{B}$}                                       & \multirow{10}{*}{2}                                                   & \multirow{10}{*}{5}                                                        & 2          & 1          & 2          & 500,000                                                          & 125,000                                                     & 1,000                                                        & 50              & 200                                                           & 83.20\%                                                      & 5.94 min                                                          & 64              & 128             & 64              & 7            \\ \cline{4-17} 
                                                                   &                                                                       &                                                                            & 2          & 1          & 2          & 500,000                                                          & 125,000                                                     & 1,000                                                        & 50              & 200                                                           & 87.20\%                                                      & 8.17 min                                                          & 128             & 256             & 128             & \textbf{8}   \\ \cline{4-17} 
                                                                   &                                                                       &                                                                            & 1          & 2          & 2          & 500,000                                                          & 125,000                                                     & 1,000                                                        & 50              & 200                                                           & 86.90\%                                                      & 5.79 min                                                          & 64              & 128             & 64              & \textbf{7}   \\ \cline{4-17} 
                                                                   &                                                                       &                                                                            & 1          & 2          & 2          & 500,000                                                          & 125,000                                                     & 1,000                                                        & 50              & 200                                                           & 86.30\%                                                      & 7.78 min                                                          & 128             & 256             & 128             & 8            \\ \cline{4-17} 
                                                                   &                                                                       &                                                                            & 2          & 2          & 1          & 500,000                                                          & 125,000                                                     & 1,000                                                        & 50              & 200                                                           & 87.10\%                                                      & 5.75 min                                                          & 64              & 128             & 64              & \textbf{7}   \\ \cline{4-17} 
                                                                   &                                                                       &                                                                            & 2          & 2          & 1          & 500,000                                                          & 125,000                                                     & 1,000                                                        & 50              & 200                                                           & 86.00\%                                                      & 7.77 min                                                          & 128             & 256             & 128             & 8            \\ \cline{4-17} 
                                                                   &                                                                       &                                                                            & 1          & 3          & 1          & 500,000                                                          & 125,000                                                     & 1,000                                                        & 50              & 200                                                           & 86.90\%                                                      & 5.7 min                                                           & 64              & 128             & 64              & 7            \\ \cline{4-17} 
                                                                   &                                                                       &                                                                            & 1          & 3          & 1          & 500,000                                                          & 125,000                                                     & 1,000                                                        & 50              & 200                                                           & 88.10\%                                                      & 7.75 min                                                          & 128             & 256             & 128             & \textbf{8}   \\ \cline{4-17} 
                                                                   &                                                                       &                                                                            & 3          & 1          & 1          & 500,000                                                          & 125,000                                                     & 1,000                                                        & 50              & 200                                                           & 87.90\%                                                      & 5.65 min                                                          & 64              & 128             & 64              & \textbf{7}   \\ \cline{4-17} 
                                                                   &                                                                       &                                                                            & 3          & 1          & 1          & 500,000                                                          & 125,000                                                     & 1,000                                                        & 50              & 200                                                           & 87.80\%                                                      & 7.97 min                                                          & 128             & 256             & 128             & \textbf{8}   \\ \hline
\multirow{18}{*}{Loop$_{B}$}                                       & \multirow{18}{*}{2}                                                   & \multirow{18}{*}{6}                                                        & 1          & 2          & 3          & 500,000                                                          & 125,000                                                     & 1,000                                                        & 50              & 200                                                           & 73.50\%                                                      & 5.81 min                                                          & 64              & 128             & 64              & \textbf{7}   \\ \cline{4-17} 
                                                                   &                                                                       &                                                                            & 1          & 2          & 3          & 500,000                                                          & 125,000                                                     & 1,000                                                        & 50              & 200                                                           & 72.70\%                                                      & 7.55 min                                                          & 128             & 256             & 128             & 8            \\ \cline{4-17} 
                                                                   &                                                                       &                                                                            & 2          & 1          & 3          & 500,000                                                          & 125,000                                                     & 1,000                                                        & 50              & 200                                                           & 75.70\%                                                      & 5.57 min                                                          & 64              & 128             & 64              & \textbf{7}   \\ \cline{4-17} 
                                                                   &                                                                       &                                                                            & 2          & 1          & 3          & 500,000                                                          & 125,000                                                     & 1,000                                                        & 50              & 200                                                           & 72.50\%                                                      & 7.58 min                                                          & 128             & 256             & 128             & 8            \\ \cline{4-17} 
                                                                   &                                                                       &                                                                            & 2          & 2          & 2          & 500,000                                                          & 125,000                                                     & 1,000                                                        & 50              & 200                                                           & 78.80\%                                                      & 5.47 min                                                          & 64              & 128             & 64              & \textbf{7}   \\ \cline{4-17} 
                                                                   &                                                                       &                                                                            & 2          & 2          & 2          & 500,000                                                          & 125,000                                                     & 1,000                                                        & 50              & 200                                                           & 78.20\%                                                      & 7.58 min                                                          & 128             & 256             & 128             & 8            \\ \cline{4-17} 
                                                                   &                                                                       &                                                                            & 3          & 1          & 2          & 500,000                                                          & 125,000                                                     & 1,000                                                        & 50              & 200                                                           & 79.30\%                                                      & 5.5 min                                                           & 64              & 128             & 64              & \textbf{7}   \\ \cline{4-17} 
                                                                   &                                                                       &                                                                            & 3          & 1          & 2          & 500,000                                                          & 125,000                                                     & 1,000                                                        & 50              & 200                                                           & 77.80\%                                                      & 7.62 min                                                          & 128             & 256             & 128             & 8            \\ \cline{4-17} 
                                                                   &                                                                       &                                                                            & 1          & 3          & 2          & 500,000                                                          & 125,000                                                     & 1,000                                                        & 50              & 200                                                           & 79.80\%                                                      & 5.75 min                                                          & 64              & 128             & 64              & 7            \\ \cline{4-17} 
                                                                   &                                                                       &                                                                            & 1          & 3          & 2          & 500,000                                                          & 125,000                                                     & 1,000                                                        & 50              & 200                                                           & 80.00\%                                                      & 7.53 min                                                          & 128             & 256             & 128             & \textbf{8}   \\ \cline{4-17} 
                                                                   &                                                                       &                                                                            & 2          & 3          & 1          & 500,000                                                          & 125,000                                                     & 1,000                                                        & 50              & 200                                                           & 84.00\%                                                      & 5.49 min                                                          & 64              & 128             & 64              & \textbf{7}   \\ \cline{4-17} 
                                                                   &                                                                       &                                                                            & 2          & 3          & 1          & 500,000                                                          & 125,000                                                     & 1,000                                                        & 50              & 200                                                           & 80.50\%                                                      & 7.54 min                                                          & 128             & 256             & 128             & \textbf{8}   \\ \cline{4-17} 
                                                                   &                                                                       &                                                                            & 3          & 2          & 1          & 500,000                                                          & 125,000                                                     & 1,000                                                        & 50              & 200                                                           & 84.20\%                                                      & 5.55 min                                                          & 64              & 128             & 64              & \textbf{7}   \\ \cline{4-17} 
                                                                   &                                                                       &                                                                            & 3          & 2          & 1          & 500,000                                                          & 125,000                                                     & 1,000                                                        & 50              & 200                                                           & 82.90\%                                                      & 7.6 min                                                           & 128             & 256             & 128             & \textbf{8}   \\ \cline{4-17} 
                                                                   &                                                                       &                                                                            & 1          & 4          & 1          & 500,000                                                          & 125,000                                                     & 1,000                                                        & 50              & 200                                                           & 85.00\%                                                      & 5.58 min                                                          & 64              & 128             & 64              & 7            \\ \cline{4-17} 
                                                                   &                                                                       &                                                                            & 1          & 4          & 1          & 500,000                                                          & 125,000                                                     & 1,000                                                        & 50              & 200                                                           & 85.30\%                                                      & 7.79 min                                                          & 128             & 256             & 128             & \textbf{8}   \\ \cline{4-17} 
                                                                   &                                                                       &                                                                            & 4          & 1          & 1          & 500,000                                                          & 125,000                                                     & 1,000                                                        & 50              & 200                                                           & 85.30\%                                                      & 5.61 min                                                          & 64              & 128             & 64              & \textbf{7}   \\ \cline{4-17} 
                                                                   &                                                                       &                                                                            & 4          & 1          & 1          & 500,000                                                          & 125,000                                                     & 1,000                                                        & 50              & 200                                                           & 83.30\%                                                      & 7.6 min                                                           & 128             & 256             & 128             & 8            \\ \hline
\multirow{4}{*}{Loop$_{C}$}                                        & \multirow{4}{*}{3}                                                    & \multirow{4}{*}{4}                                                         & 1          & 2          & 1          & 500,000                                                          & 125,000                                                     & 1,000                                                        & 50              & 200                                                           & 86.00\%                                                      & 6.3 min                                                           & 64              & 128             & 64              & \textbf{7}   \\ \cline{4-17} 
                                                                   &                                                                       &                                                                            & 1          & 2          & 1          & 500,000                                                          & 125,000                                                     & 1,000                                                        & 50              & 200                                                           & 85.90\%                                                      & 7.99 min                                                          & 128             & 256             & 128             & \textbf{8}   \\ \cline{4-17} 
                                                                   &                                                                       &                                                                            & 2          & 1          & 1          & 500,000                                                          & 125,000                                                     & 1,000                                                        & 50              & 200                                                           & 87.70\%                                                      & 5.62 min                                                          & 64              & 128             & 64              & \textbf{7}   \\ \cline{4-17} 
                                                                   &                                                                       &                                                                            & 2          & 1          & 1          & 500,000                                                          & 125,000                                                     & 1,000                                                        & 50              & 200                                                           & 86.50\%                                                      & 7.76 min                                                          & 128             & 256             & 128             & 8            \\ \hline
\multirow{11}{*}{Loop$_{C}$}                                       & \multirow{11}{*}{3}                                                   & \multirow{11}{*}{5}                                                        & 2          & 1          & 2          & 600,000                                                          & 150,000                                                     & 1,000                                                        & 50              & 200                                                           & 80.10\%                                                      & 6.84 min                                                          & 64              & 128             & 64              & \textbf{7}   \\ \cline{4-17} 
                                                                   &                                                                       &                                                                            & 2          & 1          & 2          & 600,000                                                          & 150,000                                                     & 1,000                                                        & 50              & 200                                                           & 79.20\%                                                      & 9.93 min                                                          & 128             & 256             & 128             & 8            \\ \cline{4-17} 
                                                                   &                                                                       &                                                                            & 2          & 1          & 2          & 600,000                                                          & 150,000                                                     & 1,000                                                        & 50              & 200                                                           & 77.30\%                                                      & 23.42 min                                                         & 256             & 512             & 256             & 9            \\ \cline{4-17} 
                                                                   &                                                                       &                                                                            & 1          & 2          & 2          & 600,000                                                          & 150,000                                                     & 1,000                                                        & 50              & 200                                                           & 79.70\%                                                      & 6.91 min                                                          & 64              & 128             & 64              & \textbf{7}   \\ \cline{4-17} 
                                                                   &                                                                       &                                                                            & 1          & 2          & 2          & 600,000                                                          & 150,000                                                     & 1,000                                                        & 50              & 200                                                           & 77.60\%                                                      & 9.29 min                                                          & 128             & 256             & 128             & 8            \\ \cline{4-17} 
                                                                   &                                                                       &                                                                            & 2          & 2          & 1          & 600,000                                                          & 150,000                                                     & 1,000                                                        & 50              & 200                                                           & 84.00\%                                                      & 6.81 min                                                          & 64              & 128             & 64              & \textbf{7}   \\ \cline{4-17} 
                                                                   &                                                                       &                                                                            & 2          & 2          & 1          & 600,000                                                          & 150,000                                                     & 1,000                                                        & 50              & 200                                                           & 82.10\%                                                      & 9.43 min                                                          & 128             & 256             & 128             & 8            \\ \cline{4-17} 
                                                                   &                                                                       &                                                                            & 1          & 3          & 1          & 600,000                                                          & 150,000                                                     & 1,000                                                        & 50              & 200                                                           & 87.30\%                                                      & 6.83 min                                                          & 64              & 128             & 64              & \textbf{7}   \\ \cline{4-17} 
                                                                   &                                                                       &                                                                            & 1          & 3          & 1          & 600,000                                                          & 150,000                                                     & 1,000                                                        & 50              & 200                                                           & 85.20\%                                                      & 9.48 min                                                          & 128             & 256             & 128             & 8            \\ \cline{4-17} 
                                                                   &                                                                       &                                                                            & 3          & 1          & 1          & 600,000                                                          & 150,000                                                     & 1,000                                                        & 50              & 200                                                           & 86.60\%                                                      & 6.77 min                                                          & 64              & 128             & 64              & \textbf{7}   \\ \cline{4-17} 
                                                                   &                                                                       &                                                                            & 3          & 1          & 1          & 600,000                                                          & 150,000                                                     & 1,000                                                        & 50              & 200                                                           & 85.30\%                                                      & 9.46 min                                                          & 128             & 256             & 128             & 8            \\ \hline
\end{tabular}
}
\end{table*}

\begin{table*}[]
\caption{MLP attacking performance against ($x,y$)-iPUFs (reproduction according to MLP configuration in~\cite{santikellur2019deep})}
\label{tab: IPUF MLP Reappearance}
\centering
\begin{tabular}{|c|c|c|c|c|c|c|c|c|c|c|c|c|}
\hline
\textbf{x} & \textbf{y} & \textbf{i} & \textbf{\begin{tabular}[c]{@{}c@{}}Training\\  CRPs\end{tabular}} & \textbf{\begin{tabular}[c]{@{}c@{}}Validation\\  CRPs\end{tabular}} & \textbf{\begin{tabular}[c]{@{}c@{}}Test\\ CRPs\end{tabular}} & \textbf{Epochs} & \textbf{\begin{tabular}[c]{@{}c@{}}batch\\ size\end{tabular}} & \textbf{\begin{tabular}[c]{@{}c@{}}Test\\ acc\end{tabular}} & \textbf{\begin{tabular}[c]{@{}c@{}}Training\\ Time\end{tabular}} & \textbf{layer1} & \textbf{layer2} & \textbf{layer3} \\ \hline
3          & 3          & 33         & 240,000                                                            & 60,000                                                               & 1,000                                                         & 100             & 1,000                                                          & 93.40\%                                                             & 2.1 min                                                          & 50              & 50              & 50              \\ \hline
3          & 3          & 33         & 240,000                                                            & 60,000                                                               & 1,000                                                         & 300             & 1,000                                                          & 95.80\%                                                      & 5.94 min                                                         & 50              & 50              & 50              \\ \hline
4          & 4          & 33         & 320,000                                                            & 80,000                                                               & 1,000                                                         & 100             & 1,000                                                          & 77.60\%                                                        & 3.1 min                                                          & 60              & 60              & 60              \\ \hline
4          & 4          & 33         & 320,000                                                            & 80,000                                                               & 1,000                                                         & 500             & 1,000                                                          & 95.50\%                                                            & 14.53 min                                                        & 60              & 60              & 60              \\ \hline
\end{tabular}
\end{table*}

\subsection{($x,y$)-iPUF and $Mn_{S_{1},S_{2},S_{3}}$-APUF}
\subsubsection{$Mn_{S_{1},S_{2},S_{3}}$-APUF}
Ebrahimabadi~\textit{et al.} showed that $M64_{32,16,8}$ is resilient to logistic regression (LR)~\cite{ruhrmair2010modeling}, SVM, CMA-ES~\cite{tobisch2015scaling} and 5-layer neural network attacks. However, our MLP attack on it shows attacking accuracy of 93.9\% with 200,000 training CPRs with about 20 minutes in Table~\ref{tab: mlp ff-puf}, which is breakable.

\observ{The claimed modeling resilience of $Mn_{S_{1},S_{2},S_{3}}$-APUF is debunked with our MLP attacks}

\subsubsection{($x,y$)-iPUFs}
We firstly reproduce the MLP attacks on ($x,y$)-iPUFs according to the MLP configuration by Santikellur \textit{et al.}~\cite{santikellur2019deep}. The results are detailed in Table \ref{tab: IPUF MLP Reappearance}. While the results of MLP attack performance using our MLP configurations are detailed in Table~\ref{tab: mlp ipuf}. According to Nguyen \textit{et al.}~\cite{nguyen2019interpose},
the resistance of $(x,y)$-$i$PUF to LR is similar to that of $(x/2 +y)$-XOR-APUF and when interpose position is in the middle of the challenge, the iPUF has the strongest resistance. Therefore, in the MLP configuration, we follow $l=\lceil (x/2 +y)\rceil $ to attack ($x,y$)-iPUFs and set $i=33$ (interpose position is in the middle). But note that this $l$ is not always optimal and needs to be increased by 1 in some cases. From Table~\ref{tab: mlp ipuf}, we can see that this MLP attack can successfully model $(3,3)$-iPUF and $(4,4)$-iPUF with accuracy higher than 95\%, which is better than the MLP attack in~\cite{santikellur2019deep} as in Table \ref{tab: IPUF MLP Reappearance}. The main reason is that the activation function we used is \textsf{tanh}, while \textsf{relu} is used in~\cite{santikellur2019deep}. The former \textsf{tanh} allows usage of negative value, benefiting network training~\cite{wisiol2021neural}.

We have also tested our MLP on larger scaled $(1,7)$-iPUF and $(5,5)$-iPUF, which are not tested in Santikellur \textit{et al.}~\cite{santikellur2019deep}. Attacking both $(1,7)$-iPUF and $(5,5)$-iPUF require much higher training CRPs and longer training time. As for $(5,5)$-iPUF, the accuracy is not stable per run. To gain better accuracy, multiple run trials can be made to only use the most accurate model. In our experiments, when we run 5 times, one trial is with the accuracy more than 95\%, while the rest 4 run are all about 75\%---note the later is still close to the reliability of the ($x,y$)-iPUF.

\observ{The MLP can successfully attack ($x,y$)-iPUF up to e.g., $(5,5)$-iPUF and $(1,7)$-iPUF}

\begin{table*}[]
\caption{MLP attacking performance against ($x,y$)-iPUFs (our MLP configuration as specified in last four columns)}
\label{tab: mlp ipuf}
\centering
\begin{tabular}{|c|c|c|c|c|c|c|c|c|c|c|c|c|c|}
\hline
\textbf{x} & \textbf{y} & \textbf{i}  & \textbf{\begin{tabular}[c]{@{}c@{}}Training\\  CRPs\end{tabular}} & \textbf{\begin{tabular}[c]{@{}c@{}}Validation\\  CRPs\end{tabular}} & \textbf{\begin{tabular}[c]{@{}c@{}}Test\\ CRPs\end{tabular}} & \textbf{Epochs} & \textbf{\begin{tabular}[c]{@{}c@{}}batch\\ size\end{tabular}} & \textbf{\begin{tabular}[c]{@{}c@{}}Test\\ acc\end{tabular}}   & \textbf{\begin{tabular}[c]{@{}c@{}}Training\\ Time\end{tabular}} & \textbf{layer1} & \textbf{layer2} & \textbf{layer3} & \textbf{$l$}   \\ \hline
3          & 3          & 33          & 240,000                                                            & 60,000                                                               & 1,000                                                         & 100             & 1,000                                                          & 96.70\%                                                             & 1.6 min                                                          & 16              & 32              & 16              & \textbf{5} \\ \hline
4          & 4          & 33          & 320,000                                                            & 80,000                                                               & 1,000                                                         & 100             & 200                                                           & 93.90\%                                                        & 5.6 min                                                          & 32              & 64              & 32              & 6          \\ \hline
4          & 4          & 33          & 320,000                                                            & 80,000                                                               & 1,000                                                         & 100             & 200                                                           & 94.50\%                                                           & 7.1 min                                                          & 64              & 128             & 64              & 7          \\ \hline
4          & 4          & 33          & 320,000                                                            & 80,000                                                               & 1,000                                                         & 100             & 1,000                                                          & 95.70\%                                                            & 3.1 min                                                          & 64              & 128             & 64              & \textbf{7} \\ \hline
5          & 5          & 33          & 1,200,000                                                           & 300,000                                                              & 1,000                                                         & 100             & 1,000                                                          & 74.30\%                                                            & 24.4 min                                                         & 128             & 256             & 128             & 8          \\ \hline
5          & 5          & 33          & 2,400,000                                                           & 600,000                                                              & 1,000                                                         & 100             & 10,000                                                         & 75.00\%                                                       & 30.7 min                                                         & 128             & 256             & 128             & 8          \\ \hline
\textbf{5} & \textbf{5} & \textbf{33} & \textbf{6,000,000}                                                  & \textbf{1,500,000}                                                    & \textbf{1,000}                                                & \textbf{200}    & \textbf{10,000}                                                & \textbf{95.30\%}                                             & \textbf{3.39 h}                                                  & \textbf{128}    & \textbf{256}    & \textbf{128}    & \textbf{8} \\
 \hline
5          & 5          & 33          & 6,000,000                                                           & 1,500,000                                                              & 1,000                                                         & 100             & 10,000                                                         & 96.3\%                                                             & 3.87 h                                                         & 256             & 512             & 256             & 9
\\ \hline
1          & 7          & 33          & 6,000,000                                                           & 1,500,000                                                             & 1,000                                                         & 100             & 10,000                                                         &73.60\%                                                                          & 1.62 h                                                                  & 128             & 256             & 128             & 8          \\ \hline
1          & 7          & 33          & 6,000,000                                                           & 1,500,000                                                             & 1,000                                                         & 100             & 10,000                                                         & 96.00\%                                                            & 3.77 h                                                           & 256             & 512             & 256             & \textbf{9} \\ \hline
\end{tabular}
\end{table*}
\begin{table*}[h]

\caption{MLP attacking performance against larger scaled CO-APUFs(our MLP configuration $l$ specified in last column)}
\label{tab: mlp xor amd oax}
\centering
\resizebox{\textwidth}{!}{
\begin{tabular}{|ccccccccccccccc|}
\hline
\multicolumn{1}{|c|}{\textbf{Loop.Config.ID}}     & \multicolumn{1}{c|}{\textbf{\begin{tabular}[c]{@{}c@{}}Loop.\\ Nums\end{tabular}}} & \multicolumn{1}{c|}{\textbf{\begin{tabular}[c]{@{}c@{}}Challenge\\ Size\\ (bits)\end{tabular}}} & \multicolumn{1}{c|}{\textbf{\begin{tabular}[c]{@{}c@{}}FF-APUF\\ Nums\\ ($x+y+z$)\end{tabular}}} & \multicolumn{1}{c|}{\textbf{x}} & \multicolumn{1}{c|}{\textbf{y}} & \multicolumn{1}{c|}{\textbf{z}} & \multicolumn{1}{c|}{\textbf{\begin{tabular}[c]{@{}c@{}}BER\\ ($\sigma_{noise}$)\\ (0.02;0.05)\end{tabular}}} & \multicolumn{1}{c|}{\textbf{Uniformity$_{1}$}} & \multicolumn{1}{c|}{\textbf{\begin{tabular}[c]{@{}c@{}}Training\\ CRPs\end{tabular}}} & \multicolumn{1}{c|}{\textbf{\begin{tabular}[c]{@{}c@{}}Validation\\ CRPs\end{tabular}}} & \multicolumn{1}{c|}{\textbf{\begin{tabular}[c]{@{}c@{}}Test\\ CRPs\end{tabular}}} & \multicolumn{1}{c|}{\textbf{\begin{tabular}[c]{@{}c@{}}Test\\ acc\end{tabular}}} & \multicolumn{1}{c|}{\textbf{\begin{tabular}[c]{@{}c@{}}Training\\ Time\end{tabular}}} & \textbf{$l$} \\ \hline
\multicolumn{15}{|c|}{\textbf{OAX-FF-APUF(Training CRPs,Val CRPs $\sigma_{noise}=0.02$)}}                                                                                                                                                                                                                                                                                                                                                                                                                                                                                                                                                                                                                                                                                                                                                                                                                                                                                                                                                                                  \\ \hline
\multicolumn{1}{|c|}{\multirow{6}{*}{Loop$_{A}$}} & \multicolumn{1}{c|}{\multirow{6}{*}{1}}                                            & \multicolumn{1}{c|}{\multirow{6}{*}{64}}                                                        & \multicolumn{1}{c|}{\multirow{6}{*}{7}}                                                          & \multicolumn{1}{c|}{0}          & \multicolumn{1}{c|}{0}          & \multicolumn{1}{c|}{7}          & \multicolumn{1}{c|}{\textbf{(0.199;0.418)}}                                                                  & \multicolumn{1}{c|}{\textbf{0.497}}            & \multicolumn{1}{c|}{\textbf{700,000}}                                                 & \multicolumn{1}{c|}{\textbf{175,000}}                                            & \multicolumn{1}{c|}{\textbf{1,000}}                                               & \multicolumn{1}{c|}{\textbf{49.60\%}}                                            & \multicolumn{1}{c|}{10.84 min}                                                        & 8            \\ \cline{5-15} 
\multicolumn{1}{|c|}{}                            & \multicolumn{1}{c|}{}                                                              & \multicolumn{1}{c|}{}                                                                           & \multicolumn{1}{c|}{}                                                                            & \multicolumn{1}{c|}{1}          & \multicolumn{1}{c|}{2}          & \multicolumn{1}{c|}{4}          & \multicolumn{1}{c|}{(0.174;0.378)}                                                                           & \multicolumn{1}{c|}{0.505}                     & \multicolumn{1}{c|}{700,000}                                                          & \multicolumn{1}{c|}{175,000}                                                     & \multicolumn{1}{c|}{1,000}                                                        & \multicolumn{1}{c|}{91.30\%}                                                     & \multicolumn{1}{c|}{11.52 min}                                                        & 8            \\ \cline{5-15} 
\multicolumn{1}{|c|}{}                            & \multicolumn{1}{c|}{}                                                              & \multicolumn{1}{c|}{}                                                                           & \multicolumn{1}{c|}{}                                                                            & \multicolumn{1}{c|}{1}          & \multicolumn{1}{c|}{3}          & \multicolumn{1}{c|}{3}          & \multicolumn{1}{c|}{(0.136;0.313)}                                                                           & \multicolumn{1}{c|}{0.508}                     & \multicolumn{1}{c|}{700,000}                                                          & \multicolumn{1}{c|}{175,000}                                                     & \multicolumn{1}{c|}{1,000}                                                        & \multicolumn{1}{c|}{95.00\%}                                                     & \multicolumn{1}{c|}{11.21 min}                                                        & 8            \\ \cline{5-15} 
\multicolumn{1}{|c|}{}                            & \multicolumn{1}{c|}{}                                                              & \multicolumn{1}{c|}{}                                                                           & \multicolumn{1}{c|}{}                                                                            & \multicolumn{1}{c|}{2}          & \multicolumn{1}{c|}{2}          & \multicolumn{1}{c|}{3}          & \multicolumn{1}{c|}{(0.148;0.328)}                                                                           & \multicolumn{1}{c|}{0.504}                     & \multicolumn{1}{c|}{700,000}                                                          & \multicolumn{1}{c|}{175,000}                                                     & \multicolumn{1}{c|}{1,000}                                                        & \multicolumn{1}{c|}{95.30\%}                                                     & \multicolumn{1}{c|}{11.94 min}                                                        & 8            \\ \cline{5-15} 
\multicolumn{1}{|c|}{}                            & \multicolumn{1}{c|}{}                                                              & \multicolumn{1}{c|}{}                                                                           & \multicolumn{1}{c|}{}                                                                            & \multicolumn{1}{c|}{2}          & \multicolumn{1}{c|}{3}          & \multicolumn{1}{c|}{2}          & \multicolumn{1}{c|}{(0.119;0.263)}                                                                           & \multicolumn{1}{c|}{0.496}                     & \multicolumn{1}{c|}{700,000}                                                          & \multicolumn{1}{c|}{175,000}                                                     & \multicolumn{1}{c|}{1,000}                                                        & \multicolumn{1}{c|}{95.80\%}                                                     & \multicolumn{1}{c|}{11.09 min}                                                        & 8            \\ \cline{5-15} 
\multicolumn{1}{|c|}{}                            & \multicolumn{1}{c|}{}                                                              & \multicolumn{1}{c|}{}                                                                           & \multicolumn{1}{c|}{}                                                                            & \multicolumn{1}{c|}{1}          & \multicolumn{1}{c|}{4}          & \multicolumn{1}{c|}{2}          & \multicolumn{1}{c|}{(0.101;0.234)}                                                                           & \multicolumn{1}{c|}{0.507}                     & \multicolumn{1}{c|}{700,000}                                                          & \multicolumn{1}{c|}{175,000}                                                     & \multicolumn{1}{c|}{1,000}                                                        & \multicolumn{1}{c|}{96.30\%}                                                     & \multicolumn{1}{c|}{11.23 min}                                                        & 8            \\ \hline
\multicolumn{1}{|c|}{Loop$_{B}$}                  & \multicolumn{1}{c|}{\textbf{2}}                                                    & \multicolumn{1}{c|}{64}                                                                         & \multicolumn{1}{c|}{7}                                                                           & \multicolumn{1}{c|}{1}          & \multicolumn{1}{c|}{2}          & \multicolumn{1}{c|}{4}          & \multicolumn{1}{c|}{(0.168;0.355)}                                                                           & \multicolumn{1}{c|}{0.501}                     & \multicolumn{1}{c|}{700,000}                                                          & \multicolumn{1}{c|}{175,000}                                                     & \multicolumn{1}{c|}{1,000}                                                        & \multicolumn{1}{c|}{\textbf{53.70\%}}                                            & \multicolumn{1}{c|}{31.79 min}                                                        & 9            \\ \hline
\multicolumn{1}{|c|}{\multirow{6}{*}{Loop$_{A}$}} & \multicolumn{1}{c|}{\multirow{6}{*}{1}}                                            & \multicolumn{1}{c|}{\multirow{6}{*}{64}}                                                        & \multicolumn{1}{c|}{\multirow{6}{*}{8}}                                                          & \multicolumn{1}{c|}{0}          & \multicolumn{1}{c|}{0}          & \multicolumn{1}{c|}{8}          & \multicolumn{1}{c|}{(0.232;0.491)}                                                                           & \multicolumn{1}{c|}{0.497}                     & \multicolumn{1}{c|}{800,000}                                                          & \multicolumn{1}{c|}{200,000}                                                     & \multicolumn{1}{c|}{1,000}                                                        & \multicolumn{1}{c|}{\textbf{49.10\%}}                                            & \multicolumn{1}{c|}{30.76 min}                                                        & 9            \\ \cline{5-15} 
\multicolumn{1}{|c|}{}                            & \multicolumn{1}{c|}{}                                                              & \multicolumn{1}{c|}{}                                                                           & \multicolumn{1}{c|}{}                                                                            & \multicolumn{1}{c|}{1}          & \multicolumn{1}{c|}{2}          & \multicolumn{1}{c|}{5}          & \multicolumn{1}{c|}{(0.208;0.444)}                                                                           & \multicolumn{1}{c|}{0.496}                     & \multicolumn{1}{c|}{800,000}                                                          & \multicolumn{1}{c|}{200,000}                                                     & \multicolumn{1}{c|}{1,000}                                                        & \multicolumn{1}{c|}{\textbf{50.07\%}}                                            & \multicolumn{1}{c|}{31.72 min}                                                        & 9            \\ \cline{5-15} 
\multicolumn{1}{|c|}{}                            & \multicolumn{1}{c|}{}                                                              & \multicolumn{1}{c|}{}                                                                           & \multicolumn{1}{c|}{}                                                                            & \multicolumn{1}{c|}{1}          & \multicolumn{1}{c|}{3}          & \multicolumn{1}{c|}{4}          & \multicolumn{1}{c|}{(0.175;0.390)}                                                                           & \multicolumn{1}{c|}{0.497}                     & \multicolumn{1}{c|}{800,000}                                                          & \multicolumn{1}{c|}{200,000}                                                     & \multicolumn{1}{c|}{1,000}                                                        & \multicolumn{1}{c|}{\textbf{51.90\%}}                                            & \multicolumn{1}{c|}{30.57 min}                                                        & 9            \\ \cline{5-15} 
\multicolumn{1}{|c|}{}                            & \multicolumn{1}{c|}{}                                                              & \multicolumn{1}{c|}{}                                                                           & \multicolumn{1}{c|}{}                                                                            & \multicolumn{1}{c|}{2}          & \multicolumn{1}{c|}{2}          & \multicolumn{1}{c|}{4}          & \multicolumn{1}{c|}{(0.175;0.396)}                                                                           & \multicolumn{1}{c|}{0.503}                     & \multicolumn{1}{c|}{800,000}                                                          & \multicolumn{1}{c|}{200,000}                                                     & \multicolumn{1}{c|}{1,000}                                                        & \multicolumn{1}{c|}{\textbf{49.80\%}}                                            & \multicolumn{1}{c|}{30.28 min}                                                        & 9            \\ \cline{5-15} 
\multicolumn{1}{|c|}{}                            & \multicolumn{1}{c|}{}                                                              & \multicolumn{1}{c|}{}                                                                           & \multicolumn{1}{c|}{}                                                                            & \multicolumn{1}{c|}{1}          & \multicolumn{1}{c|}{4}          & \multicolumn{1}{c|}{3}          & \multicolumn{1}{c|}{(0.140;0.312)}                                                                           & \multicolumn{1}{c|}{0.496}                     & \multicolumn{1}{c|}{800,000}                                                          & \multicolumn{1}{c|}{200,000}                                                     & \multicolumn{1}{c|}{1,000}                                                        & \multicolumn{1}{c|}{93.00\%}                                                     & \multicolumn{1}{c|}{30.22 min}                                                        & 9            \\ \cline{5-15} 
\multicolumn{1}{|c|}{}                            & \multicolumn{1}{c|}{}                                                              & \multicolumn{1}{c|}{}                                                                           & \multicolumn{1}{c|}{}                                                                            & \multicolumn{1}{c|}{2}          & \multicolumn{1}{c|}{3}          & \multicolumn{1}{c|}{3}          & \multicolumn{1}{c|}{(0.150;0.320)}                                                                           & \multicolumn{1}{c|}{0.510}                     & \multicolumn{1}{c|}{800,000}                                                          & \multicolumn{1}{c|}{200,000}                                                     & \multicolumn{1}{c|}{1,000}                                                        & \multicolumn{1}{c|}{94.60\%}                                                     & \multicolumn{1}{c|}{30.46 min}                                                        & 9            \\ \hline
\multicolumn{1}{|c|}{Loop$_{B}$}                  & \multicolumn{1}{c|}{2}                                                             & \multicolumn{1}{c|}{64}                                                                         & \multicolumn{1}{c|}{8}                                                                           & \multicolumn{1}{c|}{1}          & \multicolumn{1}{c|}{2}          & \multicolumn{1}{c|}{5}          & \multicolumn{1}{c|}{(0.185;0.403)}                                                                           & \multicolumn{1}{c|}{0.502}                     & \multicolumn{1}{c|}{800,000}                                                          & \multicolumn{1}{c|}{200,000}                                                     & \multicolumn{1}{c|}{1,000}                                                        & \multicolumn{1}{c|}{\textbf{50.10\%}}                                            & \multicolumn{1}{c|}{30.51 min}                                                        & 9            \\ \hline
\multicolumn{1}{|c|}{\multirow{4}{*}{15→80}}      & \multicolumn{1}{c|}{\multirow{4}{*}{1}}                                            & \multicolumn{1}{c|}{\multirow{4}{*}{128}}                                                       & \multicolumn{1}{c|}{\multirow{2}{*}{5}}                                                          & \multicolumn{1}{c|}{0}          & \multicolumn{1}{c|}{0}          & \multicolumn{1}{c|}{5}          & \multicolumn{1}{c|}{(0.187;0.407)}                                                                           & \multicolumn{1}{c|}{0.502}                     & \multicolumn{1}{c|}{400,000}                                                          & \multicolumn{1}{c|}{100,000}                                                     & \multicolumn{1}{c|}{1,000}                                                        & \multicolumn{1}{c|}{\textbf{50.20\%}}                                            & \multicolumn{1}{c|}{7.13 min}                                                         & 8            \\ \cline{5-15} 
\multicolumn{1}{|c|}{}                            & \multicolumn{1}{c|}{}                                                              & \multicolumn{1}{c|}{}                                                                           & \multicolumn{1}{c|}{}                                                                            & \multicolumn{1}{c|}{1}          & \multicolumn{1}{c|}{2}          & \multicolumn{1}{c|}{2}          & \multicolumn{1}{c|}{(0.144;0.331)}                                                                           & \multicolumn{1}{c|}{0.502}                     & \multicolumn{1}{c|}{400,000}                                                          & \multicolumn{1}{c|}{100,000}                                                     & \multicolumn{1}{c|}{1,000}                                                        & \multicolumn{1}{c|}{\textbf{49.40\%}}                                            & \multicolumn{1}{c|}{5.26 min}                                                         & 7            \\ \cline{4-15} 
\multicolumn{1}{|c|}{}                            & \multicolumn{1}{c|}{}                                                              & \multicolumn{1}{c|}{}                                                                           & \multicolumn{1}{c|}{6}                                                                           & \multicolumn{1}{c|}{1}          & \multicolumn{1}{c|}{2}          & \multicolumn{1}{c|}{3}          & \multicolumn{1}{c|}{(0.189;0.425)}                                                                           & \multicolumn{1}{c|}{0.506}                     & \multicolumn{1}{c|}{500,000}                                                          & \multicolumn{1}{c|}{125,000}                                                     & \multicolumn{1}{c|}{1,000}                                                        & \multicolumn{1}{c|}{\textbf{50.00\%}}                                            & \multicolumn{1}{c|}{8.78 min}                                                         & 8            \\ \cline{4-15} 
\multicolumn{1}{|c|}{}                            & \multicolumn{1}{c|}{}                                                              & \multicolumn{1}{c|}{}                                                                           & \multicolumn{1}{c|}{7}                                                                           & \multicolumn{1}{c|}{1}          & \multicolumn{1}{c|}{2}          & \multicolumn{1}{c|}{4}          & \multicolumn{1}{c|}{(0.217;0.466)}                                                                           & \multicolumn{1}{c|}{0.495}                     & \multicolumn{1}{c|}{700,000}                                                          & \multicolumn{1}{c|}{175,000}                                                     & \multicolumn{1}{c|}{1,000}                                                        & \multicolumn{1}{c|}{\textbf{49.30\%}}                                            & \multicolumn{1}{c|}{12.58 min}                                                        & 8            \\ \hline
\multicolumn{15}{|c|}{\textbf{FF-APUF(Training CRPs,Val CRPs $\sigma_{noise}=0.05$)}}                                                                                                                                                                                                                                                                                                                                                                                                                                                                                                                                                                                                                                                                                                                                                                                                                                                                                                                                                                                      \\ \hline
\multicolumn{1}{|c|}{15→80,85,90,95,100}          & \multicolumn{1}{c|}{5}                                                             & \multicolumn{1}{c|}{128}                                                                        & \multicolumn{1}{c|}{1}                                                                           & \multicolumn{1}{c|}{0}          & \multicolumn{1}{c|}{0}          & \multicolumn{1}{c|}{1}          & \multicolumn{1}{c|}{(/;0.070)}                                                                               & \multicolumn{1}{c|}{0.480}                     & \multicolumn{1}{c|}{200,000}                                                          & \multicolumn{1}{c|}{5,000}                                                       & \multicolumn{1}{c|}{1,000}                                                        & \multicolumn{1}{c|}{\textbf{66.30\%}}                                            & \multicolumn{1}{c|}{2.07 min}                                                         & 6            \\ \hline
\multicolumn{1}{|c|}{15→80,85,90;95,100,105}      & \multicolumn{1}{c|}{6}                                                             & \multicolumn{1}{c|}{128}                                                                        & \multicolumn{1}{c|}{1}                                                                           & \multicolumn{1}{c|}{0}          & \multicolumn{1}{c|}{0}          & \multicolumn{1}{c|}{1}          & \multicolumn{1}{c|}{(/;0.062)}                                                                               & \multicolumn{1}{c|}{0.502}                     & \multicolumn{1}{c|}{200,000}                                                          & \multicolumn{1}{c|}{5,000}                                                       & \multicolumn{1}{c|}{1,000}                                                        & \multicolumn{1}{c|}{\textbf{85.40\%}}                                            & \multicolumn{1}{c|}{2.54 min}                                                         & 7            \\ \hline
\end{tabular}
}
\end{table*}
\vspace{-0.2cm}

\section{Discussion}\label{sec:discussion}
\subsection{Side Channel Information}
As mentioned above, the side channel information such as power~\cite{ruhrmair2014efficient}, timing~\cite{ruhrmair2014efficient}, photonic information~\cite{tajik2017photonic}, response reliability information~\cite{becker2015gap} or Hamming weight~\cite{becker2015pitfalls} has been utilized to assist modeling attacks, in particular, against APUFs and XOR-APUFs. However, the power, timing, and photonic side channel information are expensive (especially the photonic) or require expertise to collect compared with the reliability information.

As reliability information has been utilized to facilitate the modeling of the XOR-APUFs~\cite{becker2015gap}, there has no explicit studies on how to efficiently use the reliability side channel information to attack CO-APUFs, especially the FF-APUFs and its variants. This provides interesting future work.

\subsection{Heterogeneous Underlying APUFs}\label{sec:hetero} 
One can employ different PUFs as underlying PUFs when constructing the composited PUFs, in particular, the CO-APUFs concerned in this work. For example, in the OAX-FF-APUF or XOR-FF-APUF, the FF-APUF can with different configurations per se but share the e.g., same length of challenge such as 64 bits. It has shown a better modeling resilience of a heterogeneous XOR-FF-APUF compared with homogeneous XOR-FF-APUF in~\cite{avvaru2020homogeneous}. Future work can evaluate the modeling resilience of CO-APUFs composited by heterogeneous PUFs using deep learning.

\subsection{Larger Scaled CO-APUFs}
Firstly, increasing the underlying PUF's scale can increase the modeling resilience (i.e. as shown in Table~\ref{tab: mlp oax-ff-puf}). We have carried further experiments on $z$-XOR-FF-APUF and ($x,y,z$)-OAX-FF-APUF by using more than six XOR-FF-APUFs, as detailed in Table~\ref{tab: mlp xor amd oax}. We can see that the modeling accuracy is close to guessing when the scale is properly increased (underlying FF-APUFs is no less than 7). Note that the reliability will be deteriorated. To make the XOR-FF-APUF to be useful, the underlying FF-APUF reliability must be firstly improved---so that we set the noise setting to be $0.02$ for showing this effect. The OAX-FF-APUF has a much wider gap between its reliability and the modeling accuracy, affirming its advantage over XOR-FF-APUF.
Secondly, this work tests the common length of challenge that is 64 bits. Increase the length of challenge is a means of further improving the CO-APUF modeling resilience. We have tested 128-bit challenge for the FF-APUF, XOR-FF-APUF and OAX-FF-APUF, as detailed in Table~\ref{tab: mlp xor amd oax}. As a comparison, the 64-stage $5$-XOR-FF-APUF is breakable with 400,000 CPRs (i.e. $Loop_A$ in Table~\ref{tab: mlp XOR-ff-puf}), but exhibits only a 50.20\% attacking accuracy under same settings except that the number of stage increases to 128. Thirdly, the loop positions and number of loops of the FF-APUF and its variants can be flexibly changed and increased, respectively, to increase the modeling resilience. For instance, if the FF-APUF has two intermediate arbiters and the ending points of the loops are far away, more CRPs is required for the successful FF-APUF attack~\cite{avvaru2019effect}. As for the ($x,y$)-iPUF, breaking the ($x,y$) to be e.g., ($1,10$) that is still non-trivial without high computations~\cite{nguyen2019interpose}. Future work can increase the scale of the CO-APUFs while optimizing their reliability as possible.

\observ{Increasing the scale (i.e. the number of underlying FF-APUFs or stage of APUFs) can greatly increase CO-APUF's modeling resilience. But this should be done by minimizing the reliability to retain the CO-APUF's practicality (i.e. utilizing the OAX-FF-APUF instead of its counterpart XOR-FF-APUF)
}

\subsection{Protected Challenge-response Interface}
Based on our evaluations on the CO-APUFs and recent study on the XOR-APUFs (i.e. \cite{wisiol2021neural}), it appears that the small scaled APUFs variants are very challenging to resist modeling attacks assisted with DL techniques when the challenge-response interface are plainly exposed. It is worth to combine other security building blocks with APUF or its variants to enable their customized secure usages, such as lightweight secure authentication. For instances, the TREVERSE exploits the Hash to protect the APUF response~\cite{gao2020treverse} and Lockdown-PUF~\cite{yu2016lockdown} limits the number of exposed APUF CRPs to prevent modeling attacks, while incurring as minimal additional overhead as possible.
we can employ not only the APUF but also the FF-APUF for re-composition, where both APUF and FF-APUF still share the same challenge.

\section{Conclusion}\label{sec:conclusion}
Five CO-APUFs are systematically evaluated under unified experimental settings for fair quantitative comparisons in terms of each CO-APUF configurations' reliability and uniformity. The OAX-APUF exhibits best reliability. All CO-APUFs except the FF-APUF have satisfactory uniformity. We have demonstrated that DL techniques are effective to model the CO-APUFs with best attacking performance provided by the MLP, as MLP is good at learning any non-linear functions. Larger scaled CO-APUFs still appears hard to learn with personal accessible computing resource (a typical personal computer used in our case), but usually they are with worsen reliability. Therefore, future strong PUF designs built upon basic PUF composition should always take the reliability optimization into consideration when injecting higher non-linearity into PUF structure---the OAX-FF-APUF is shown to be a promising alternative. Considering the MLP outperforms (attacking with shorter period or/and training CRPs) previous widely used machine learning techniques such as LR and CMA-ES, it is highly recommended to always leverage it to examine newly devised strong PUF candidates.


\bibliographystyle{IEEEtran}
\bibliography{IEEEToCArxiv}

\end{document}